\documentclass[aps,showpacs,showkeys,nofootinbib,preprint]{revtex4}
\usepackage{epsfig,amssymb,bm,graphicx,color,amsmath,rotating}
\usepackage{xcolor}
\usepackage[percent]{overpic}
\usepackage{mwe,xcolor}
\usepackage{transparent}

\begin{document} {\normalsize }
\interfootnotelinepenalty=10000

\title{Impact of laser polarization on
q-exponential photon tails in non-linear Compton scattering}
%
\author{B.~K\"ampfer}
\affiliation{Helmholtz-Zentrum  Dresden-Rossendorf, 01314 Dresden, Germany}
\affiliation{Institut f\"ur Theoretische Physik, TU~Dresden, 01062 Dresden, Germany}

\author{A.~I.~Titov}
\affiliation{Bogoliubov Laboratory of Theoretical Physics, JINR, Dubna 141980, Russia}
 
\begin{abstract}
Non-linear Compton scattering of ultra-relativistic electrons traversing 
high-intensity laser pulses generates also hard photons.
These photon high-energy tails are considered for
parameters in reach at the forthcoming experiments LUXE and E-320. 
We consider the invariant differential cross sections
$d \sigma / du$ between the IR and UV regions and analyze the impact of
the laser polarization and find q-deformed exponential shapes. 
(The variable $u$ is the light-cone momentum-transfer from initial electron
to final photon.)
Optical laser pulses of various durations are compared with the monochromatic
laser beam model which uncovers the laser intensity parameter in the range
$\xi = 1 \cdots 10$. Some supplementary information is provided
for the azimuthal final-electron/photon distributions and the photon
energy-differential cross sections.
\end{abstract}



\maketitle

\section{Introduction}

The planned experiments LUXE at DESY  
\cite{Abramowicz:2019gvx,Abramowicz:2021zja,Altarelli:2019zea,Hartin:2018sha}
and E-320  at  FACET-II \cite{E_320} aim at studying
fundamental QED processes within strong laser fields characterized
by intensities in the order of $10^{20}$~W/cm${^2}$. A particular feature
is the use of a high-quality electron ($e^-$, mass $m$, charge $- \vert e \vert$) 
beam provided by an accelerator, thus 
possessing fairly well controlled parameters. The available beams uncover 
ultra-relativistic energies
$E_{e^-} =  10 \cdots 50$~GeV. Even for non-ultra-strong lasers
in the so-called transition region in between weak-field and strong-field limits
of QED, 
the field strength, which the electron experiences in its local rest system,
reaches values in the order of the so-called (critical)
Sauter-Schwinger electric field 
${\cal E}_{crit} = m^2 / \vert e \vert \approx 1.3 \times 10^{18}$
V/m,\footnote{
{\color{black}
Natural units with $\hbar = c = 1$ are used.}}
thus enabling a test of strong-field QED in a hitherto less explored regime
and continuing the seminal experiments 
\cite{Bula:1996st,Burke:1997ew,Bamber:1999zt,Poder:2018ifi,Cole:2017zca}
towards the precision regime. 
{\color{black} For an introduction to the state-of-the-art
physics case of strong-field QED
and a deeper survey on quantum processes in strong e.m.\ fields,
we refer the interested reader to the theory sections in 
\cite{Abramowicz:2021zja} together with 
\cite{Abramowicz:2019gvx,Altarelli:2019zea,Hartin:2018sha,E_320,Turcu:2016dxm,Meuren:2020nbw}.}

{\color{black}
When considering electron-laser interactions,
e.g.\ the nonlinear Compton scattering as
the Furry picture process
$e_L (p) \to e_L (p') + \gamma (k')$, where $e_L$ stands for the laser-dressed
electron and $\gamma$ for the emitted photon,
two invariant parameters are often used in plane-wave backgrounds
to characterize the entrance channel \cite{Ritus,DiPiazza:2011tq}:
\begin{eqnarray}
  \xi &=& \frac{e a}{m}, \label{I1}\\
  \chi &=& \xi \frac{k \cdot p}{m^2},
  \label{I2}
\end{eqnarray}
where $a$ is the amplitude of electromagnetic potential}
and momentum four-vectors $k$, $k'$ and $p$
refer to the laser-beam wave-vector, the $out$-photon 
(with energy $\omega'$),  and the $in$-electron, respectively.
The $out$-electron four-momentum is $p'$.
The  invariant  laser intensity parameter 
in the lab.\ system reads
$\xi = (m/\omega) ({\cal E}/{\cal E}_{crit})$.
{\color{black} The meaning of the parameter $\chi$ becomes more transparent
in the electron's rest system:
\begin{eqnarray}
  \chi = \frac{\cal E}{{\cal E}_{crit}}
  \left( \frac{E_{e^-}}{m} + \sqrt{\frac{E_{e^-}^2}{m^2} - 1}\right)
\bigg\rvert_{E_{e^-} = m}
  = \frac{\cal E}{{\cal E}_{crit}}~.
  \label{I3}
\end{eqnarray} 
The frequency of a Ti:Sapphire laser is 
$\omega  =  1.55$~eV, thus $m/\omega \gg 1$.} 
In other words,
even for lasers with intensities $\xi \gtrsim 1$, 
i.e.\ ${\cal E}/{\cal E}_{crit} \ll 1$ in the lab.,
the Lorentz boost of the electric field strength ${\cal E}$
lets the quantum parameter become $\chi \lesssim 1$, thus testing the
sub-critical up to the critical regime, ${\cal E} \lesssim {\cal E}_{crit}$ for 
$E_{e^-} = 17.5$~GeV in the electron's rest system.

Among the options at LUXE and E-320
are investigations of non-linear effects in Compton scattering, Breit-Wheeler
pair production and trident pair production. Here, we consider non-linear
Compton scattering as one-photon emission by an electron traversing a
laser pulse. 
We focus on the photon tails: 
the region beyond the Klein-Nishina edge, i.e.\ excluding the
IR region, and prior to the kinematic limit, i.e.\ excluding the UV region
towards the kinematical limit. The considered
laser intensities $\xi \sim 1$ are in between weak-field and strong-field
limits, where approximation schemes are often hardly universally applicable.
({\color{black}
Among important approximation schemes are the locally constant field approximation
\cite{Harvey:2014qla} and improvements 
\cite{Ilderton:2018nws,DiPiazza:2017raw,DiPiazza:2018bfu}} and
the locally monochromatic approximation scheme \cite{Heinzl:2020ynb} as well.)

It was already noted by Ritus \cite{Ritus} that non-linear Compton scattering
for circular laser polarization is fundamentally different from linear polarization. 
One may expect this: The classical trajectory of a point-like charge in
a circularly polarized e.m.\ wave is essentially on a circle {\it perpendicular}
to the wave vector $\vec k$, while in a linearly polarized e.m.\ wave it
is on the figure-8 curve {\it in direction} of $\vec k$
and parallel to the polarization
vector $\vec a$. Correspondingly, the radiation patterns of the moving charges
are expected to differ. In fact, in a monochromatic plane wave, the circular
polarization facilitates the dead cone
effect~\cite{Titov:2019kdk,Harvey:2009ry,Maltoni:2016ays},
i.e.\ all harmonics beyond the
first one are zero for on-axis back-scattering,
while for the linear polarization only the even harmonics are zero, 
{\color{black} which is a clear manifestation
of specific properties of transition matrix elements described 
by their distinct basic functions.} 

We are going to compare the photon tails in non-asymptotic regions
of $\xi$ and $u = k \cdot k' / (k \cdot p - k \cdot k')$ 
for circular and linear polarizations.
Such comparative studies appeared already in the recent literature, 
e.g.\ \cite{Titov:2019kdk,Heinzl:2020ynb},
but not with emphasis on the intermediate photon tails. 
The tails in the considered region display a q-deformed exponential
shape of the invariant cross sections $d \sigma / du$, which we
quantify accordingly.
We also show that the monochromatic laser model is a useful reference,
supported by numerical results of laser pulses when considering the
differential spectra $d \sigma / du$ or $d \sigma / d \omega'$.

Our note is organized as follows. In section \ref{basics} we recall the
basic formulas of one-photon emission by electrons in laser pulses and
in a monochromatic laser beam and in a constant cross field; 
we also supply certain limits of these formulas. 
The central part, section \ref{results}, is devoted to
the numerical evaluation of these less transparent formulas. 
There, we also comment on the azimuthal distribution 
$d^2 \sigma / du \, d \phi_{e'}$
of the recoil electron described by the four-momentum $p'$. 
In section \ref{qexp}, we describe 
the adjustment of q-exponentials to the $u$-differential cross sections. 
The discussion in section \ref{discussion}
is devoted to integrated cross sections with cut-off, some remarks on emissivity
of thermalized systems and the relation of the spectra  
$d \sigma / du$ vs.\ $d \sigma / d \omega'$.
We conclude in section \ref{summary}.

\section{Basics}\label{basics}

The here considered process of non-linear Compton scattering
(cf.~\cite{DiPiazza:2020wxp,Seipt:2020diz,Valialshchikov:2020dhq} 
for recent developments and detailed citations)
is the Furry-picture one-photon emission by an electron
traversing an external electromagnetic field
which approximates the laser on different levels of sophistication.
This section recaps the used formulas in the subsequent numerical analysis.

\subsection{Laser pulses}

The laser pulse model for plane waves is described by the four-potential
in axial gauge, $A^{(i)} = (0, \vec A^{(i)})$, with
\begin{equation}\label{laser}
\vec A^{(i)} = f(\phi) \left(\vec a_x(i) \cos \phi + \vec a_y(i) \sin \phi \right)
\end{equation}
where $\vec a_x(circ)^2 = \vec a_y(circ)^2 = \vec a_x(lin)^2= m^2\xi^2/e^2$,
and $\vec a_y(lin)^2=0$;
the polarization vectors $\vec a_x(circ)$ and  $\vec a_y(circ)$ are mutually
orthogonal. 
{\color{black} The intensity parameter $\xi$ is considered henceforth as
independent invariant variable characterizing the laser. The mean energy densities
$\propto \langle A^2 \rangle$ differ therefore by a factor of two
for circular and linear polarizations at given value of $\xi$.}  
We ignore a possible non-zero value
of the carrier envelope phase and focus on symmetric envelope functions
$f(\phi)$ w.r.t.\ the invariant phase $\phi = k \cdot x$. 

We recall the formalism of \cite{Titov:2019kdk} to display the relevant equations
for the calculation of the differential cross sections
for circular ($i = circ$) and linear ($i = lin$) polarizations:
\begin{equation} \label{sigma}
\frac{d \sigma^{(i)}}{d u} = \frac{\alpha^2}{\xi^2N^{(i)}} \frac{1}{k \cdot p}
\frac{1}{(1+u)^2} \int_0^{2 \pi} d \phi_{e'} \int_{\ell_{\rm min}}^\infty d \ell \,
w^{(i)}(\ell) ,
\end{equation}
where $\ell_{\rm min}=um^2/2(k\cdot p)$ and
\begin{equation}
w^{(i)}(\ell) = \left\{
\begin{array}{l}
-2 \vert \tilde Y_\ell \vert^2 \, + \, \xi^2 \left(1+\frac{u^2}{2(1+u)} \right)
\left( \vert Y_{\ell - 1}\vert^2 + \vert Y_{\ell +1} \vert^2
- 2 \mbox{Re} \tilde Y_\ell X_\ell^*) \right) 
\, \mbox{for} \, i = circ,\\
-2\vert \tilde A_0 \vert^2 + 2\xi^2 \left(1 + \frac{u^2}{2(1 + u)} \right)
\left( \vert \tilde A_1\vert ^2 - \mbox{Re}\tilde A_0 \tilde A_2^* \right) 
\, \mbox{for} 
\, i = lin ~.
\end{array}
\right.
\end{equation}

The Lorentz and gauge invariant quantity squared $\xi^2$ 
is the classical non-linearity parameter characterizing solely the laser beam,
and $\alpha$ stands for the fine-structure constant.  
The above defined invariant $u = k \cdot k' / k \cdot p'$ is used 
to characterize the $out$-photon.
The normalization factors $N^{(i)}$, which are related to the
average density of the e.m.\ field $\langle \cal{E} \rangle$,
are expressed through the envelope functions as
\begin{equation}
N^{(i)}_0 = \frac{1}{2 \pi}\int\limits_{-\infty}^{\infty}
d\phi \left(f^2(\phi)+{f'}^2(\phi)\right) \times
\left\{
\begin{array}{l}
1  \, \mbox{for} \, i = circ,\\
\cos^2 (\phi)\ \, \mbox{for} \, i = lin,\\
\end{array}
\right.
\end{equation}
with the asymptotic values 
$N^{(circ)}_0 \simeq \Delta/\pi$ and
and $N^{(lin)}_0 \simeq \Delta/2\pi$ at $\Delta/\pi \gg1$.
The functions $Y_\ell, X_\ell$ and $\tilde Y_\ell $ are defined by
\begin{eqnarray}
Y_\ell(z)&=&\frac{1}{2\pi} {\rm e}^{-i\ell(\phi_{e'})}\int\limits_{-\infty}^{\infty}\,
d\phi\,{f}(\phi)
\,{\rm e}^{i\ell\phi-i{\cal P}^{(circ)}(\phi)} ~,\nonumber\\
X_\ell(z)&=&\frac{1}{2\pi}{\rm e}^{-i\ell(\phi_{e'})} \int\limits_{-\infty}^{\infty}\,
d\phi\,{f^2}(\phi)
\,{\rm e}^{i{\ell} \phi-i{\cal P}^{(circ)}(\phi)}~,\nonumber\\
\widetilde Y_\ell(z) &=& \frac{z}{2l} 
\left[Y_{\ell+1}(z) + Y_{\ell-1}(z)\right] - \xi^2\frac{u}{u_l}\,X_\ell(z)~,
\label{YX1}
\end{eqnarray}
where 
$\phi_{e'}$ is the azimuthal angle of the $out$-electron 
$u_\ell={2\ell(k \cdot p)}/{m^2}$ and
 \begin{eqnarray}
{\cal P}^{(circ)}(\phi)&=&z\int\limits_{-\infty}^{\phi}\,d\phi'\,f(\phi')
\cos(\phi'-\phi_{e'}) 
 - \frac{\xi^2m^2u}{2(k\cdot p)}\int\limits_{-\infty}^\phi\,d\phi'\,f^2(\phi')~.
\label{YX2}
\end{eqnarray}
The functions $\widetilde A_m(\ell)$ for $m=1,2$ read
 \begin{eqnarray}
 \widetilde A_m(\ell)
 =\frac{1}{2\pi}\int\limits_{-\infty}^{\infty}d\phi\,f^m(\phi)
 \cos^m(\phi)\,{\rm e}^{i\ell\phi
 -i{\cal P}^{(lin)}(\phi)}~,
 \label{A1}
 \end{eqnarray}
\begin{eqnarray}
{\cal P}^{(lin)}(\phi) &=&
\tilde\alpha(\phi)- \tilde\beta(\phi)~,  \label{II5}\\
\tilde\alpha(\phi)&=&\hat \alpha\int\limits_{-\infty}^{\phi}d\phi'
f(\phi')\cos(\phi')~, \quad \hat \alpha = z \cos\phi_{e'} \label{II55}\\
\tilde\beta(\phi)&=&4 \hat \beta\int\limits_{-\infty}^{\phi}d\phi'
f^2(\phi')\cos^2(\phi')~, 
\quad
\hat \beta=\frac{u \xi^2m^2}{8 k \cdot p}~,
\label{A2}
\end{eqnarray} 
and the function $\widetilde A_0(\ell)$ follows from the identity~\cite{Titov:2019kdk}
\begin{eqnarray}
\ell\widetilde A_0(\ell) -  \hat \alpha \widetilde A_1(\ell)
+ 4 \hat \beta \widetilde A_2(\ell) =0~.
\label{A3}
\end{eqnarray}
All basis functions have the arguments
$z = z_\ell={2\ell \xi}\sqrt{\frac{u}{u_\ell}(1-\frac{u}{u_\ell})}$
and are defined for $0\le u\le u_\ell$ vanishing elsewhere.
Note the correspondence with the analog expressions for the
monochromatic model below, where the discrete harmonic number $n$
appears instead of the internal continuous variable $\ell$.

\subsection{Special: monochromatic laser beam model}\label{mono_model}
 
A monochromatic laser field in plane-wave approximation
is described by Eq.~(\ref{laser}) with $f(\phi) = 1$. The
invariant differential cross sections for one-photon emission
read \cite{Ritus}:
\begin{equation} \label{sigma_IPA}
\frac{d \sigma^{(i)}_{IPA}}{du} = \frac{\alpha^2}{\xi^2} 
\frac{1}{(k \cdot p)\,N^{(i)}}
\frac{1}{(1+u)^2}
\sum_{n = 1}^\infty \int_0^{2 \pi} d \phi_{e'} \, F_n^{(i)}(z_n), 
\end{equation}
where $N^{(circ)}=1$ and $N^{(lin)}=\frac12$ and
\begin{equation} \label{F_IPA}
F_n^{(i)} = \left\{
\begin{array}{l}
 - 2 J_n(z_n)^2 + \xi^2
\left(1 + \frac{u^2}{2(1+u)} \right) 
\left(
J_{n+1}(z_n)^2 + J_{n-1}(z_n)^2 - 2 J_n(z_n)^2 \right) 
\quad \mbox{for} \, i = circ,\\
 -2 A_0^2 + 2\xi^2 
\left(1 + \frac{u^2}{2(1 + u)} \right)
(A_1^2 - A_0 A_2)  \quad \mbox{for} \, i = lin,
\end{array}
\right.
\end{equation}
for $0 \le u \le u_n$ and $F_n^{(i)} = 0$ elsewhere.
$J_n$ are Bessel function of the first kind (independent of the $out$-electron
azimuthal angle $\phi_{e'}$),
and the functions $A_m (z_n)$, $m \in 0, 1, 2$, are defined by
\begin{equation}
A_m(n, \hat \alpha, \hat \beta) = \frac{1}{2 \pi} \int_{- \pi}^\pi d \phi
\cos^m (\phi) \cos (n \phi - \hat \alpha \sin \phi + \hat \beta \sin 2 \phi ),
\end{equation}
where $\hat \alpha = z_n \cos \phi_{e'}$ and 
$\hat \beta = \frac{ \xi^2 u m^2}{8 k \cdot p}$. 
The arguments are 
$z_n(u, u_n) = \frac{2 n \xi}{\sqrt{1 + c_i\xi^2}} 
\sqrt{\frac{u}{u_n} (1 - \frac{u}{u_n})}$
with $u_n = \frac{2 n k \cdot p}{m^2 (1 +c_i\xi^2)}$
with $c_{circ}=1$ and $c_{lin}=\frac12$.
The effective masses $m^2 (1 + c_i\xi^2)$ and their role in the
(quasi-) momentum balance as well as the relation to asymptotic four-momenta
($p/p'$ for $in/out$-electrons and $k/k'$ for $in/out$-photons) are discussed in detail
in \cite{LL,Harvey:2009ry}.
The differential non-linear Compton cross section 
for circular polarization has been used fairly often
as standard reference \cite{LL,Harvey:2009ry}. We use the label IPA
as acronym of ``infinite pulse approximation".

The large-$\xi$ limit of Eq.~(\ref{sigma_IPA}) reads
(cf.\ \cite{Ritus})
\begin{eqnarray} \label{sigma_largexi}
\frac{d\sigma^{(i)}_{IPA, large-\xi}}{du}&=&
\frac{8\alpha^2}{\pi^{b_{(i)}}\,\xi^2(k\cdot p)\,N^{(i)}} 
\frac{1}{(1+u)^2} 
\int\limits_{0}^{\pi} d \psi
\int\limits_{-\infty}^{\infty} d\tau \nonumber\\ 
& \times &
v_{(i)}^{\frac13}
[-\Phi^2(y_{(i)}) + v_{(i)}^{-\frac23}
\left(1+\frac{u^2}{2(1+u)}\right)
(y\Phi^2(y_{(i)}) +{\Phi'}^2(y_{(i)}))]
\end{eqnarray}
with Airy function $\Phi$ and its derivative $\Phi'$ and
$y_{(i)} = ( 1 + \tau^2) v_{(i)}^{\frac23}$ and $v_{(i)} = u/(2 \chi S^{(i)})$,
where 
\begin{equation}
S^{(i)} = \left\{
\begin{array}{l}
1 \quad \mbox{for} \, i = circ,\\
\sin \psi \quad \mbox{for} \, i = lin,
\end{array} 
\quad \quad
\right.
b^{(i)} = \left\{
\begin{array}{l}
1 \quad \mbox{for} \, i = circ,\\
2 \quad \mbox{for} \, i = lin.
\end{array} 
\right.
\end{equation}

The expressions for circular and linear polarizations look similar.
The principle difference is that the mod-square of the matrix element, 
in the case of circular polarization, does not depend on the azimuthal 
angle of the outgoing particle. This leads to a one-dimensional integral,
in contrast to a two-dimensional integral over auxiliary variables $\psi$ and $\tau$  
in the case of linear polarization.
Formally, the corresponding cases are related by $lin \to circ$ via 
\begin{equation}
N^{lin}\to N^{circ},\qquad
\psi=\frac{\pi}{2},\qquad
\int\limits_0^\pi d \psi \to \pi~.
\label{L10}
\end{equation}

\subsection{Constant cross field}

The asymptotic cross section for circular polarization 
in the large-$\xi$ limit, $d\sigma^{(circ)}_{IPA, large-\xi}/du$,
Eq.~(\ref{sigma_largexi}),
coincides with the one-photon emission in a constant cross field (ccf)
\cite{Ritus}:
\begin{eqnarray}
\frac{d\sigma_{ccf}}{du}
&=&-\frac{4 \pi \alpha^2}{m^2 \xi} \frac{1}{\chi} \frac{1}{(1 + u)^2}
\left(
\int_{z}^{\infty} dy \, \Phi(y) + \frac{2}{z} 
\left[1 + \frac{u^2 }{2 (1 + u)} \right]
\Phi'(z) \right)
\label{ccf1}\\
\stackrel{u \gg 1, \chi}{\longrightarrow} 
&\simeq&
\frac{2 \sqrt{\pi} \alpha^2}{m^2 \xi}
\chi^{- 1/2} u^{- 3/2}
\exp \left(- \frac{2 u}{3 \chi} \right),
\label{ccf2}
\end{eqnarray}
where the last line is the large-$u$ approximation,
denoted hereafter as $d \sigma_{ccf, large-u} / du$.
$\Phi(z)$ and $\Phi(z)'$ stand again
for the Airy function and its derivative
with arguments $z = (u / \chi)^{2/3}$.

\section{Numerical results}\label{results}

The following numerical results are for parameters motivated by
LUXE  \cite{Abramowicz:2019gvx,Abramowicz:2021zja}:
$E_{e^-} =17.5$~GeV and $\omega = 1.55$~eV  
for the idealized case of a head-on collision, meaning 
$k \cdot p / m^2 = 0.2078$ in the entrance channel.
{\color{black} To specify the laser model (\ref{laser}),
we employ $f(\phi) = 1 / \cosh(\phi/\pi N)$,
where $N$ characterizes the number of oscillations in the pulse.
The typical pulse duration of the envisaged experiments
is $T \approx 30$~fs which corresponds to the number of cycles in a pulse
$N \approx T \omega /(2\pi) \approx 10$.
In view of the general physical interest and
for methodological purposes,
we extended our consideration to the region of ultra-short
(sub-cycle) and short pulses with $1 \le N \le 10$.}
{\color{black} Results for circular and linear polarizations are
compared at given value of $\xi$.}\footnote{{\color{black} One could equally well
compare circular and linear polarizations at given laser intensity,
which would mean to chose $\xi$ for circular polarization and 
$\xi \sqrt{2}$ for linear polarization. This would not change our conclusions.}}  

\subsection{Invariant cross sections {\boldmath $d \sigma / du$}}

\begin{figure}[tb!]
\includegraphics[width=0.49\columnwidth]{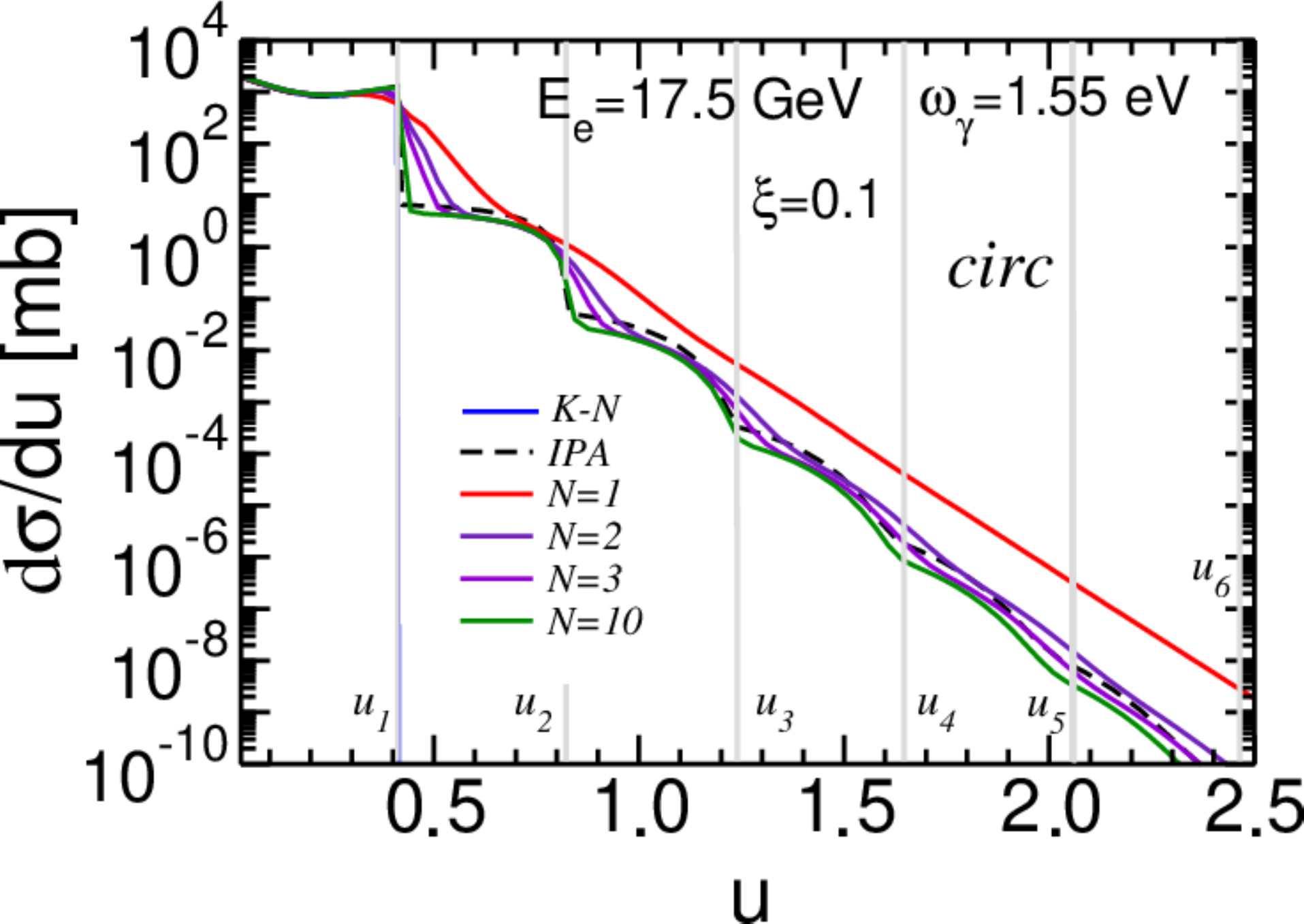}
\includegraphics[width=0.49\columnwidth]{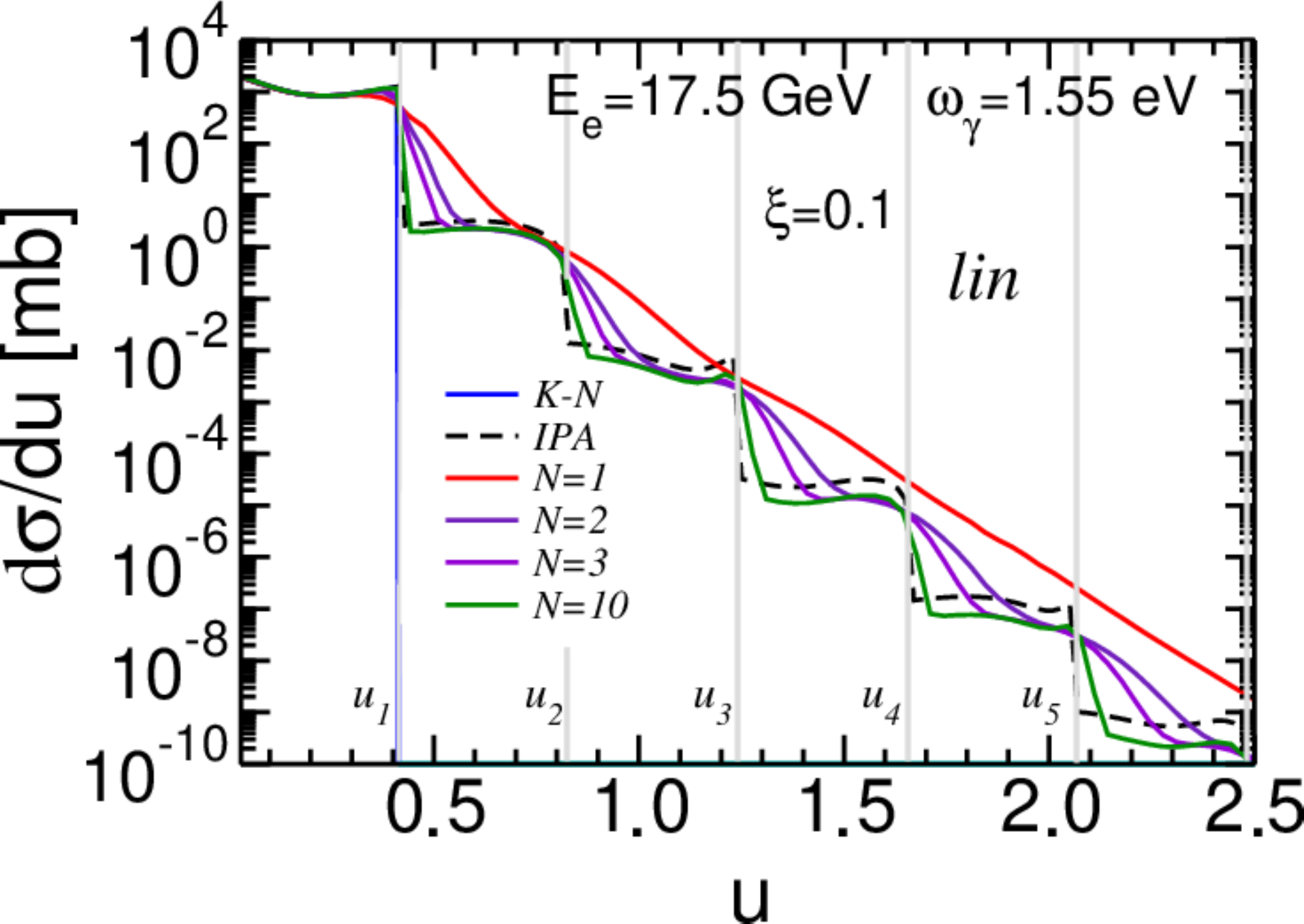}   
\includegraphics[width=0.49\columnwidth]{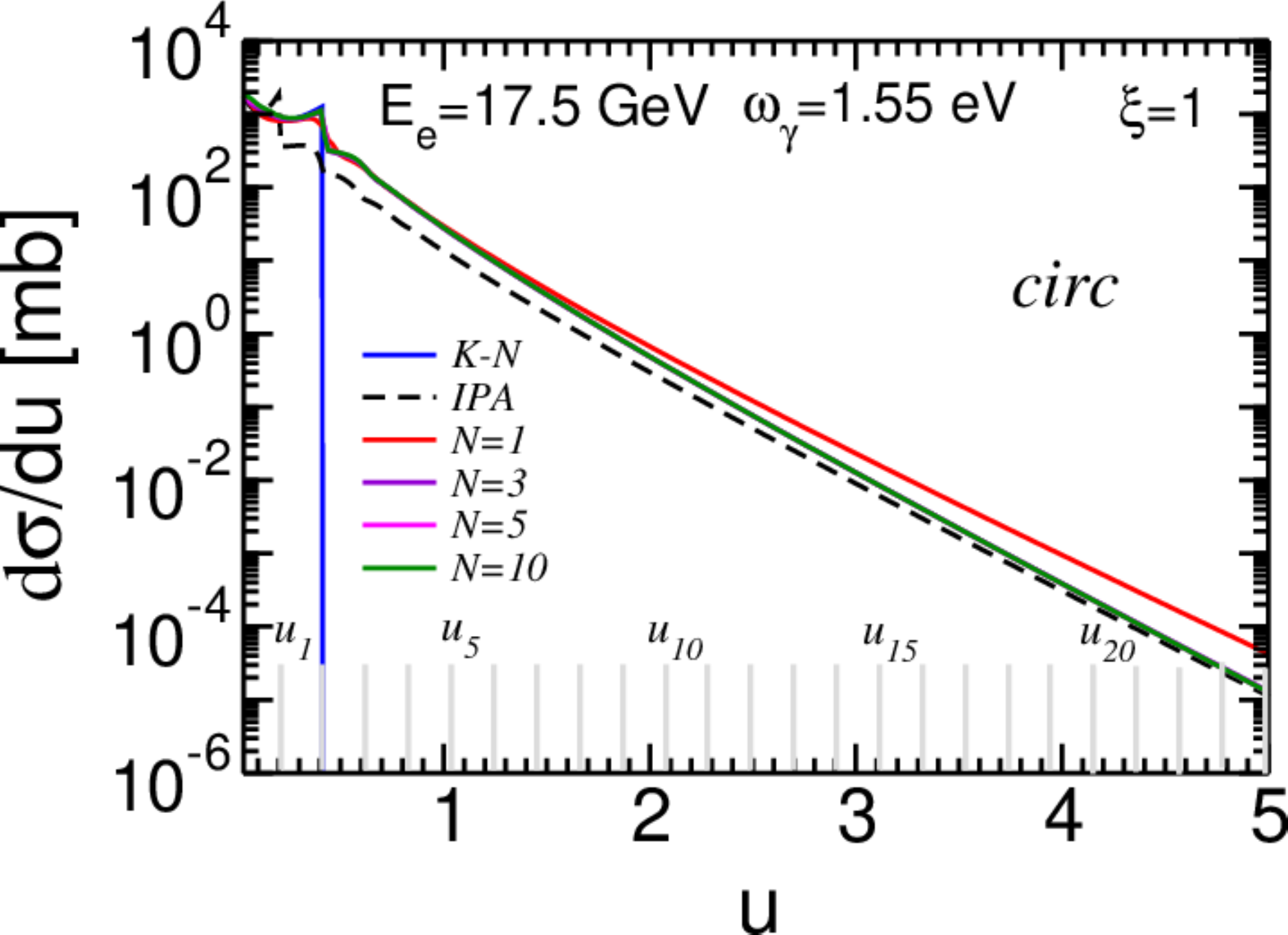}
\includegraphics[width=0.49\columnwidth]{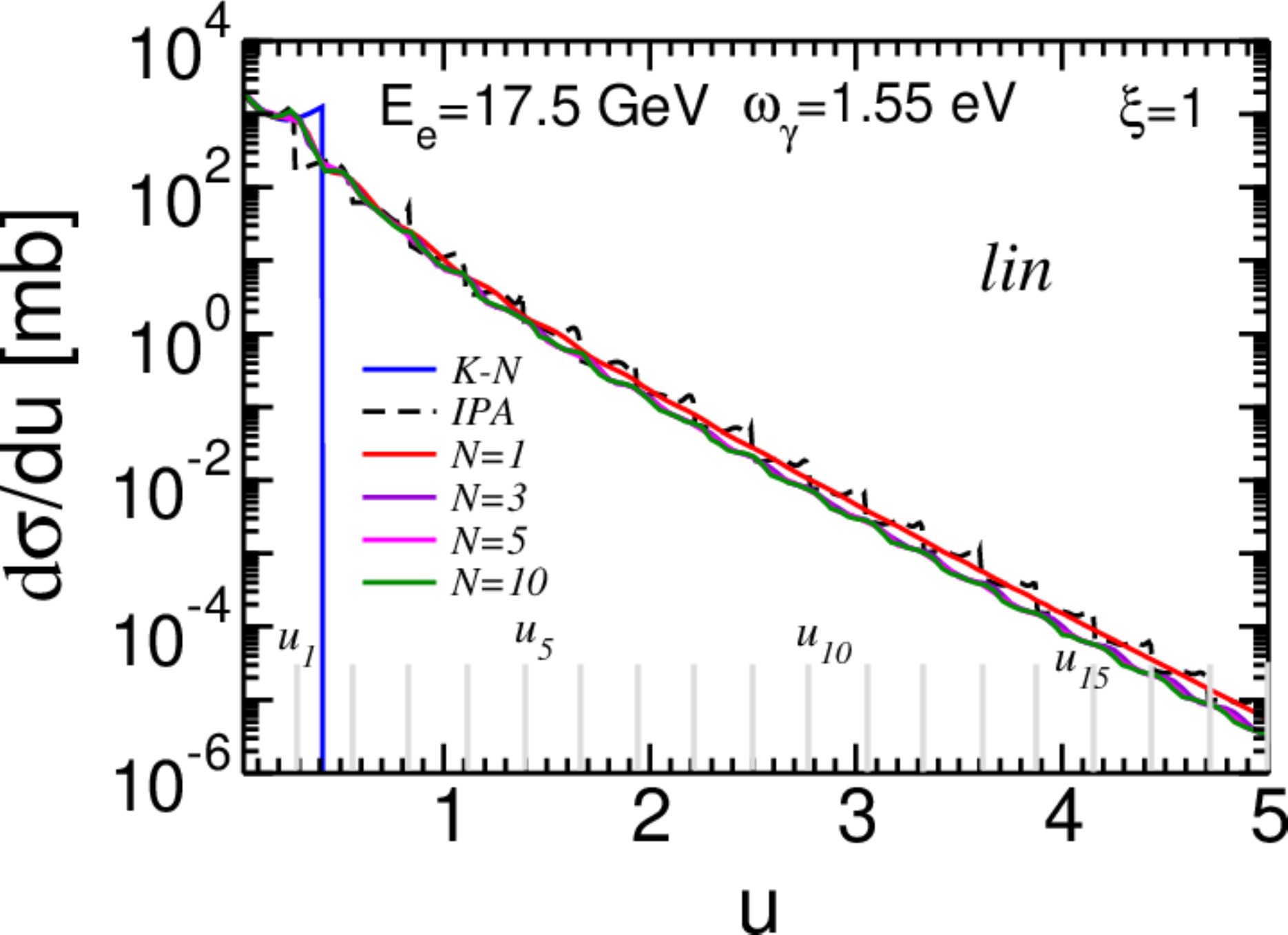}
\caption{Invariant differential cross sections $d \sigma^{(i)} / du$ 
for $\xi =0.1$ (top panels) and
for $\xi =1$ (bottom panels). 
The solid curves are for ultra-short and short laser pulses with
envelope function $f(\phi) = 1 / \cosh(\phi/(\pi N))$ for various values of 
$N = 1, 3, 5$ and 10 as given by the legends
(the $N=10$ curves are on top of the $N= 3$, 5 curves in the used scale).
The dashed curves are for the monochromatic laser beam model
$d \sigma^{(i)}_{IPA} / du$
(\ref{sigma_IPA}).
The solid blue curve depicts the Klein-Nishina (K-N) cross section
$d \sigma_{IPA} / du \vert_{\xi \to 0}$.
\color{black}
The gray vertical grid lines in top panels and
gray combs in bottom panels display the IPA harmonic positions $u_n$. 
\color{black}
Left panels: circular polarization, right panels: linear polarization. 
\label{dsigma/du}}
\end{figure}

Let us consider the above pulse envelope function 
$f(\phi) = 1 / \cosh(\phi/(\pi N))$
to elucidate the impact of a finite pulse duration and contrast it later on
with the monochromatic laser beam model and some approximations
thereof.
Differential spectra $d \sigma / du$ are exhibited in Fig.~\ref{dsigma/du}.
The panels in the top row are for a low field intensity, $\xi=0.1$.
In this case, our model manifests a significant sensitivity of cross sections
to the pulse duration parameterized by $N$. Most notable is the
dependence on the laser polarization: 
For circular polarization (left top panel), the
harmonic structures, which arise when crossing the respective upper
limit $u_{\ell, n}$ of a certain harmonic, are rather modest, while
for linear polarization (right top panel)
they persist in a much more pronounced manner
up to larger values of the variable $u$. The monochromatic model
(\ref{sigma_IPA}) 
{\color{black} (note some even-odd modification of peaked structures
to smooth ones at the respective harmonic thresholds $u_n$ for
linear polarization,\footnote{
Such an even-odd change has been reported already in \cite{Ivanov:2004fi},
see figure 6 there.}
in particular for $\xi = 0.1$)}
reproduces the pulse model (\ref{sigma}) fairly well
for longer pulses.
{\color{black}
 Note that, in the partial probabilities of a sub-cycle pulse with $N = 1$
  or smaller, an effective high-energy component  in the Fourier spectrum
of the laser pulse is generated,
  which leads to a significant increase of the corresponding
  cross sections/probabilities.
  When the duration of a pulse increases this effect decreases and the
  enhancement vanishes~\cite{Titov:2012rd}.
One can see a qualitative agreement between results 
for infinite pulses 
  and finite pulses with $N=10$. Some visible difference
  between them is explained
  by the quite different basic functions, say Bessel functions for IPA  and
  functions (\ref{YX1}), for circular polarization. This difference is much
  smaller than the scale of variation of the cross sections in the range 
$0<u<5$ which is many orders of magnitude.}

Ultra-short pulses, e.g.\ $N = 1$ exhibit hardly
the harmonic structures, both for circular and linear polarizations,
in particular at larger values of $u$.
Such a pulse duration dependence fades away considerably
at the higher intensity $\xi=1$ exhibited in the bottom panels.
Focusing first on circular polarization (see left bottom panel) and
the region $u > 0.5$, one observes a structureless and smooth spectrum
with a tiny dependence on the pulse duration for $N > 1$. Only the
ultra-short pulse result with $N = 1$ is lifted at large values of $u$. 
Taking the
monochromatic model as reference, one recognizes the approach of the
$N> 1$ results to it at large values of $u$.
Only at $u \lesssim 3$, the monochromatic
model falls somewhat short (a factor up to about three) 
in relation to the pulse results.
The difference in cross sections $d \sigma^{(i)} / du$
with $N=3\cdots10$
is comparable to the line thickness of the curves and is not visible
at the given scale.

The dependence on the pulse duration 
for linear polarization is also weak 
(see right bottom panel of Fig.~\ref{dsigma/du}).
The short pulses, $N = 3$,  5 and 10,
carry a weak remainder of the harmonic structures up to large
values of $u$. For the ultra-short pulse, $N = 1$, the spectrum is completely 
smooth beyond the Klein-Nishina edge, similar to the circularly polarized
laser pulse. The spectra are somewhat steeper than the ones for circular
polarization.
The monochromatic model, in contrast,  
displays pronounced harmonic structures up to large values of $u$. 
Remarkably, in the range of our interest, $u \in [0.5, 4]$, 
the pulse model results
are represented nicely by the smoothed monochromatic model. 
We conclude that, for the gross features, 
the monochromatic model provides a good guidance
for the tails of the spectra $d \sigma^{(i)} / du$ 
beyond the stark harmonic structures
at small values of $u$, which extend roughly up to the Klein-Nishina edge.
The occurrence of the photon tails beyond the Klein-Nishina edge,
$u > u_{K-N} = 0.416$, is a clear
signature of the multi-photon effects, becoming operative in intense lasers,
both for circular and linear polarizations.

\begin{figure}[tb!] 
\includegraphics[width=0.49\columnwidth]{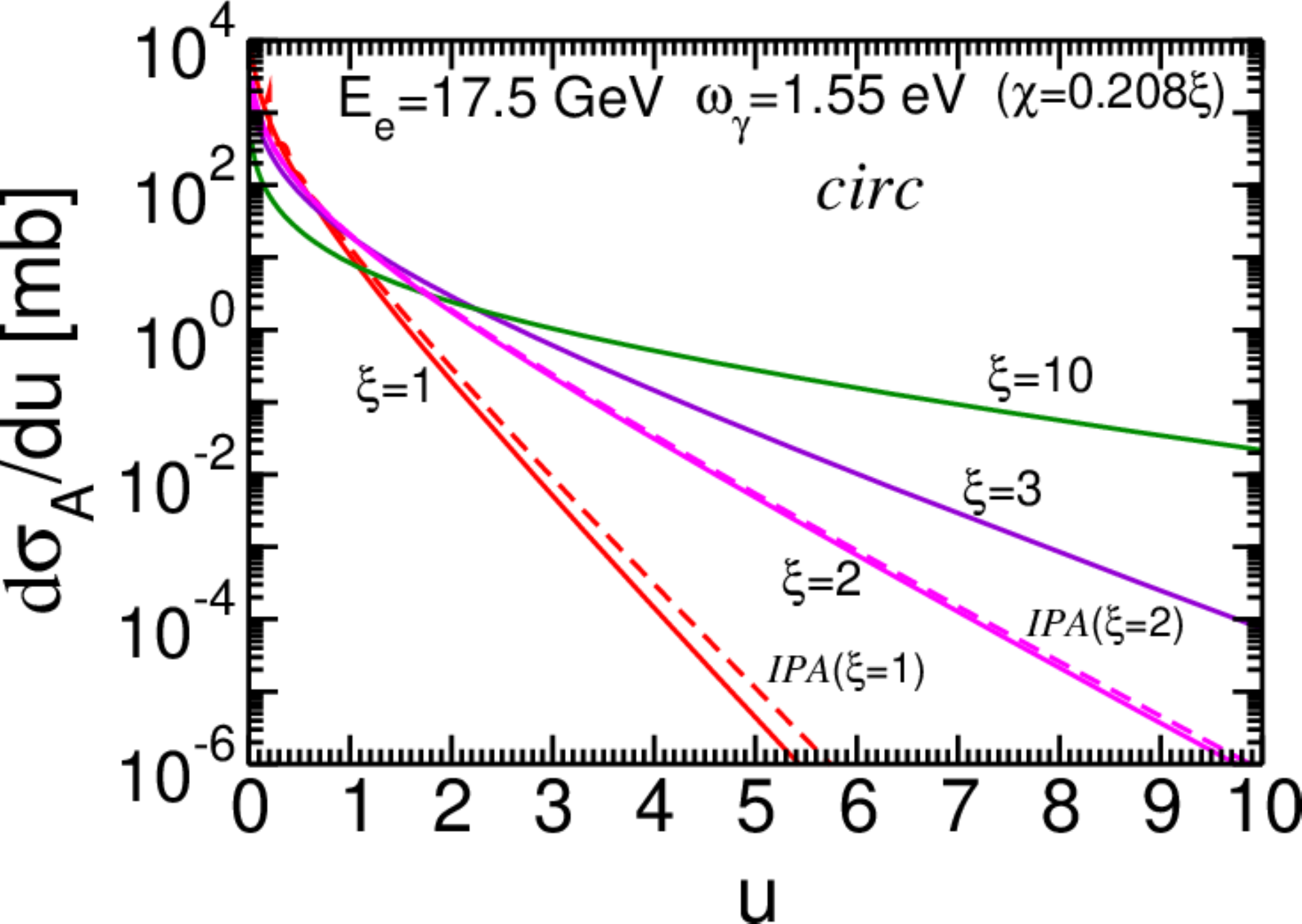}
\includegraphics[width=0.49\columnwidth]{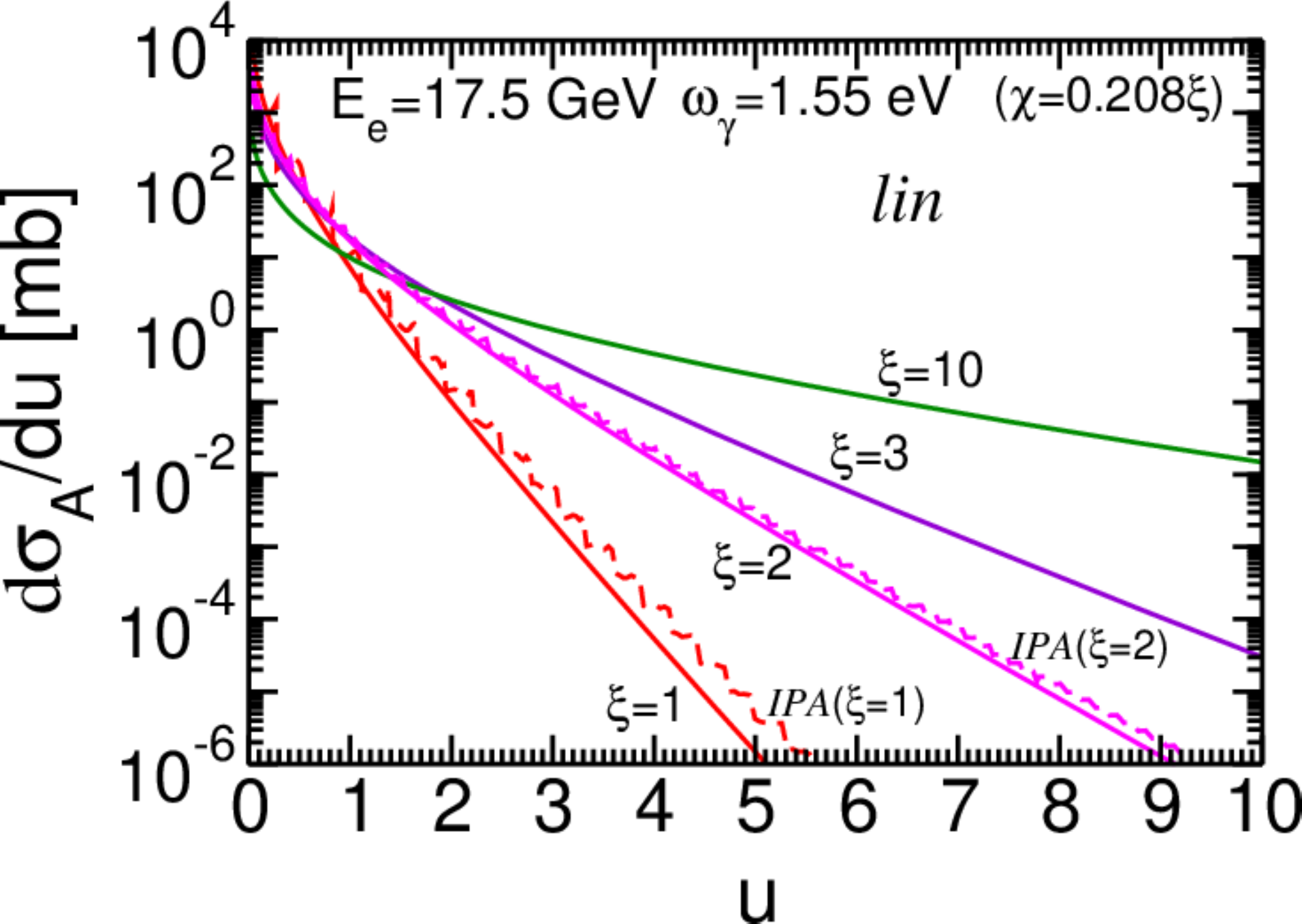}
  \caption{{\color{black}
      Invariant differential cross sections, based on an expression
    derived for asymptotically large field intensity
    $d\sigma^{(i)}_A/du\equiv d\sigma^{(i)}_{IPA, large-\xi} / du$,} 
Eq.~(\ref{sigma_largexi}),
for $\xi = 1, 2, 3, 10$ (solid curves).
For a comparison, the dashed curves with labels IPA
($\xi = 1$ and 2)
are for the monochromatic laser beam model
$d \sigma^{(i)}_{IPA}/du$, Eq.~(\ref{sigma_IPA}).
Left panel: circular polarization, right panel: linear polarization. 
\label{dsigma/du_Largexi}}
\end{figure}

Given the proximity of the spectra for laser pulses with the monochromatic
model, we consider now the change of the spectral shapes with increasing
values of the laser intensity $\xi$. We employ the large-$\xi$ approximation
Eq.~(\ref{sigma_largexi}). As seen in Fig.~\ref{dsigma/du_Largexi},
this approximation is useful already for $\xi = 2$ and not too bad for
$\xi = 1$. Of course, the harmonic structures for linear polarization
are not captured by Eq.~(\ref{sigma_largexi}), which is not problematic
when considering the gross features of the spectra beyond the Klein-Nishina
edge.   
Interestingly, the harmonic structures for the case of linear polarization
persist from the small-$u$ region up to large values of $u$ for $\xi \lesssim 2$,
see right panel of Fig.~\ref{dsigma/du_Largexi}.
The harmonic structures fade away for $\xi > 2$. The overall
pattern resembles on first sight the one for the above circular polarization case.

\begin{figure}[tb!] 
\includegraphics[width=0.49\columnwidth]{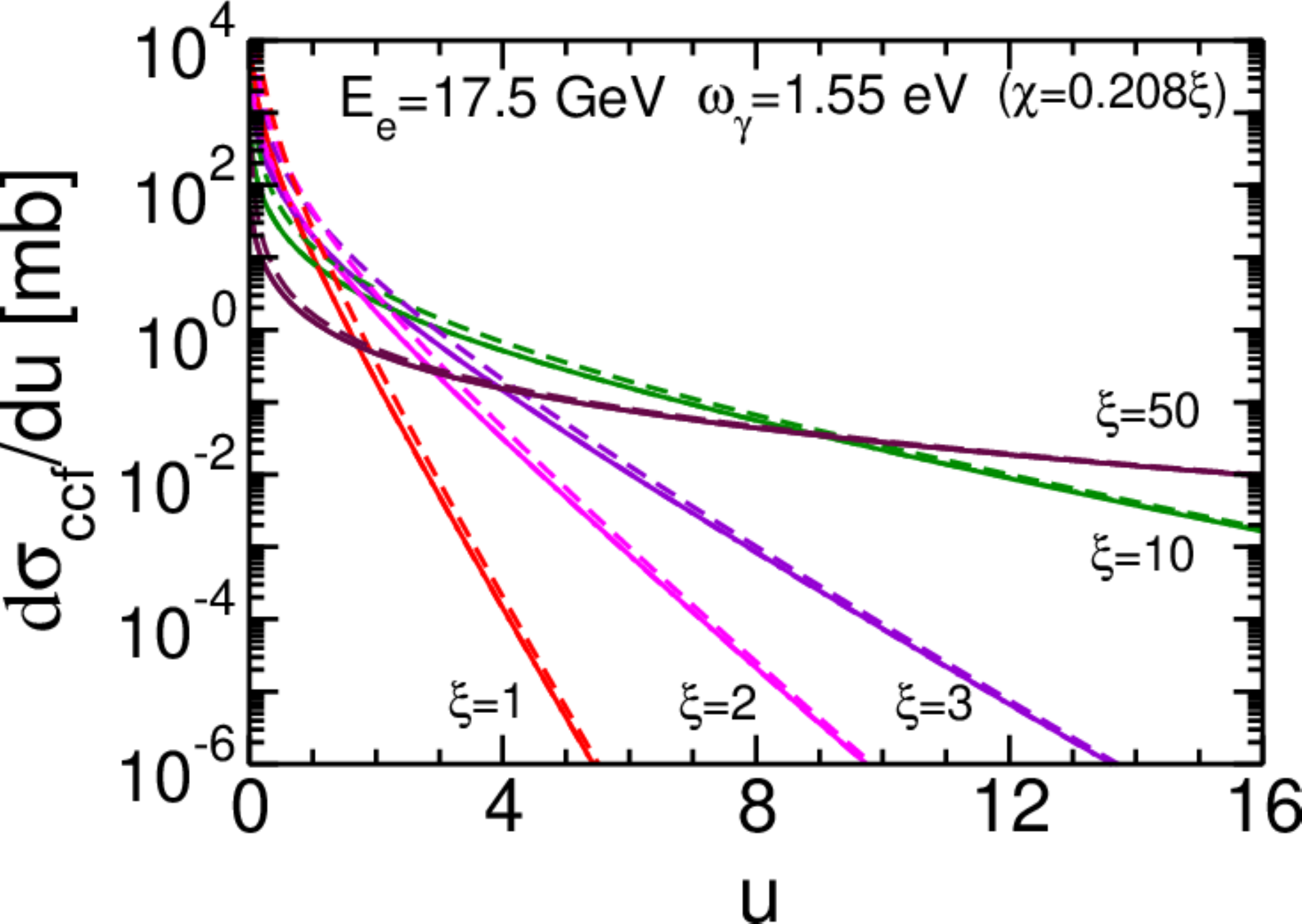}
\includegraphics[width=0.47\columnwidth]{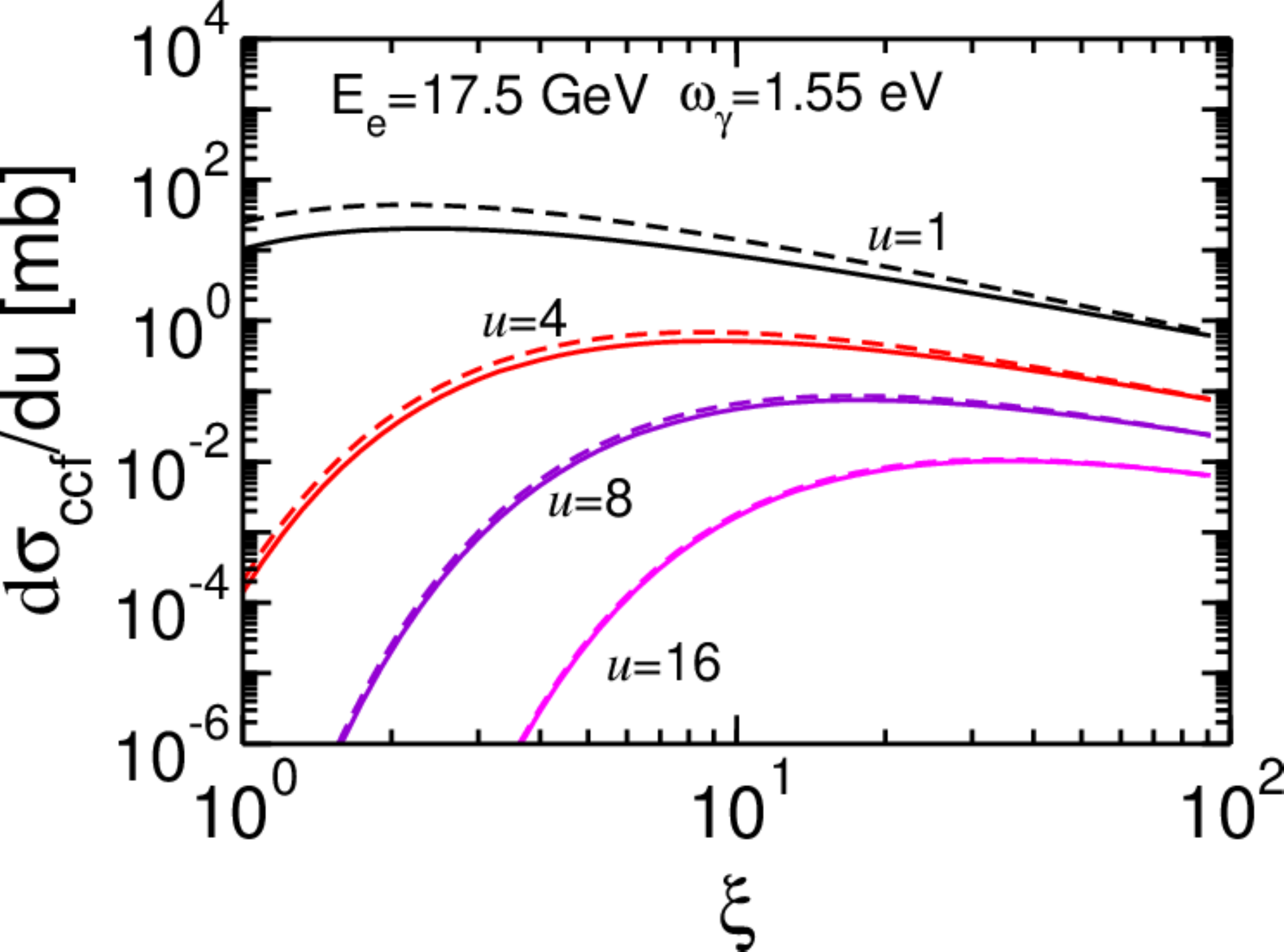}
\caption{Comparison of constant cross field results of 
$d \sigma_{ccf} / du$, Eq.~(\ref{ccf1}), (solid curves) and
$d \sigma_{ccf, large-u} / du$, Eq.~(\ref{ccf2}), (dashed curves).
Note that the large-$\xi$
approximation $d \sigma^{(circ)}_{IPA, large-\xi} / du$,
Eq.~(\ref{sigma_largexi}), coincides with $d \sigma_{ccf} / du$.
Left panel: These cross sections as a function of $u$
for $\xi = 1, 2, 3, 10$ and 50.
Right panel: These cross sections as a function of $\xi$ for 
$u =  1,$ 4, 8 and 16.
\label{ccf-predictions}}
\end{figure}

The results for $\xi=1,\,2,\,3,\,10$ and 50 based on
Eqs.~(\ref{sigma_largexi}, \ref{ccf1}) and (\ref{ccf2})  
are exhibited in Fig.~\ref{ccf-predictions}.
The solid and dashed curves
are for Eqs.~(\ref{sigma_largexi}), which is the same as (\ref{ccf1}),
and (\ref{ccf2}), respectively.
The cross section (\ref{ccf2})
with the simple exponential shape $\exp(-\frac23u/\chi)$
modified by the pre-exponential factor $\chi^{-1/2} \, u^{-3/2}$
is close to the exact result in a wide region of $\xi$ and $u$
and may be used for estimates. 

\subsection{Azimuthal electron distributions 
{\boldmath $d^2 \sigma / du \, d\phi_{e'}$}}

\begin{figure}[tb!] 
\includegraphics[width=0.49\columnwidth]{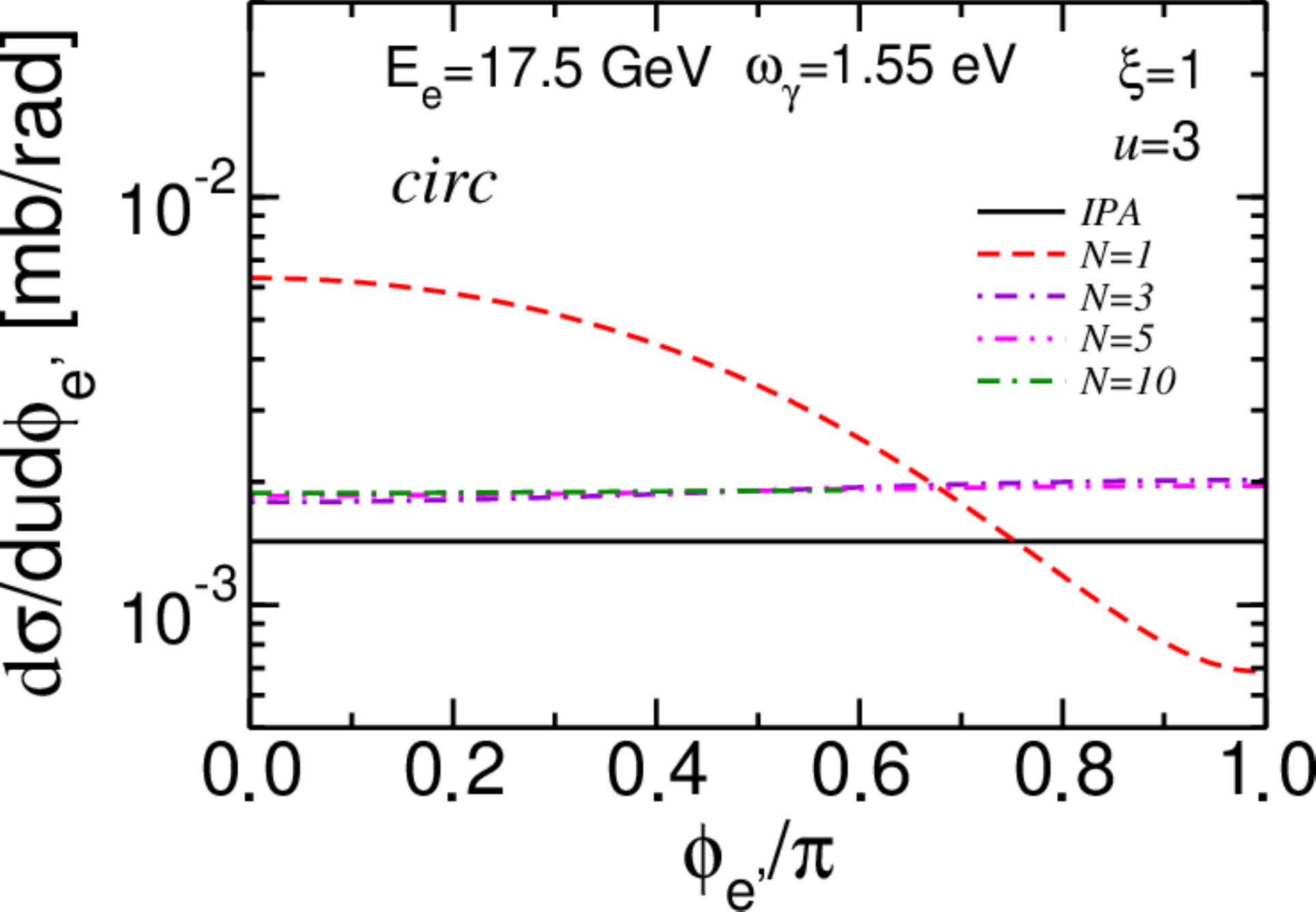}
\includegraphics[width=0.49\columnwidth]{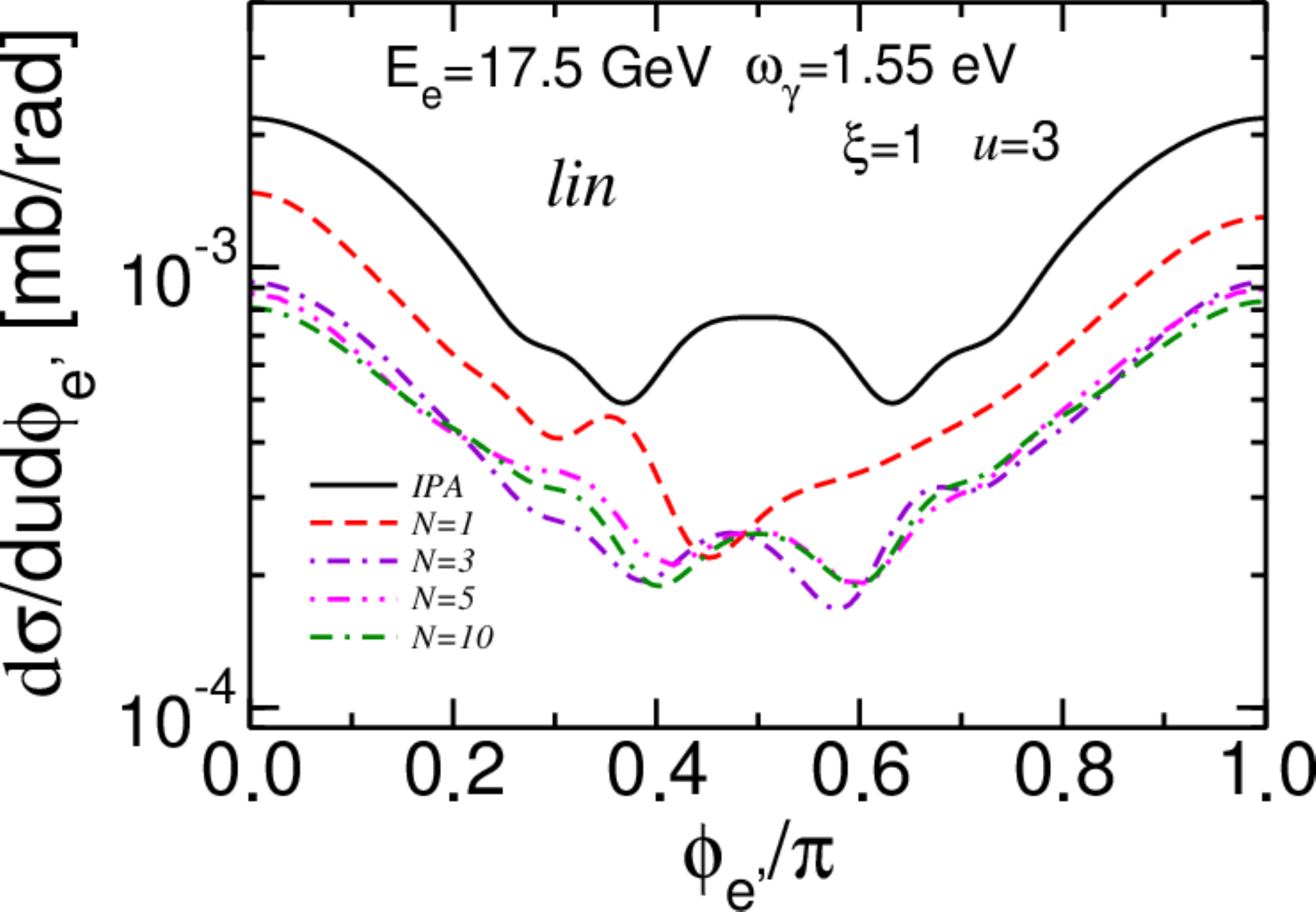}
\put(-271,134){{\small u=3}}
\caption{Invariant double-differential cross sections
$d^2 \sigma^{(i)} / d u \, d \phi_{e'}\vert_{u = 3}$ 
for various values of $N$ as given in the legends (colored curves).
The monochromatic model, $d^2 \sigma^{(i)}_{IPA} / du \, d \phi_{e'}$,
is depicted by solid black curves.
Left panel: circular polarization, right panel: linear polarization,
both for $\xi = 1$.
\label{azi}}
\end{figure}

After this comparison of the invariant differential cross section
$d \sigma / du$ with a chain of approximations,
$d \sigma_{IPA} /du$, $d \sigma_{IPA, large-\xi} /du$,
$d \sigma_{ccf} /du$ and $d \sigma_{ccf, large-u} /du$,
let us turn, as an aside, to an invariant double-differential cross section
with respect to some azimuthal dependence. 
The azimuthal $out$-electron distributions
{\color{black} is known \cite{Seipt:2013taa,Seipt:2016rtk,Blackburn:2020jaz}
}
to carry also imprints of the laser polarization. 
This is evidenced in Fig.~\ref{azi}, 
where the double differential cross sections
$d^2 \sigma^{(i)} / du \, d \phi_{e'}$ are exhibited at $u = 3$. 
For circular polarization (see left panel), short pulses characterized
by $N \ge 2$ facilitate a near-flat distribution. Only the ultra-short
pulse, $N = 1$, displays a pronounced non-uniform distribution. By definition,
the distribution for the monochromatic laser model, 
$d^2 \sigma^{(circ)}_{IPA}/du \, d \phi_{e'}$, is completely flat.
In contrast, the case of linear polarization exhibits clearly non-uniform distributions
(see right panel). The monochromatic model is symmetric around
$\pi/2$, with main maxima at $\phi_{e'} = 0$ and $\pi$.
The symmetry around $\pi/2$ gets more and more lost for shorter pulses, 
$N < 10$, with completely asymmetric distribution for the ultra-short
pulse, $N = 1$. 

\begin{figure}[tb!] 
\includegraphics[width=0.49\columnwidth]{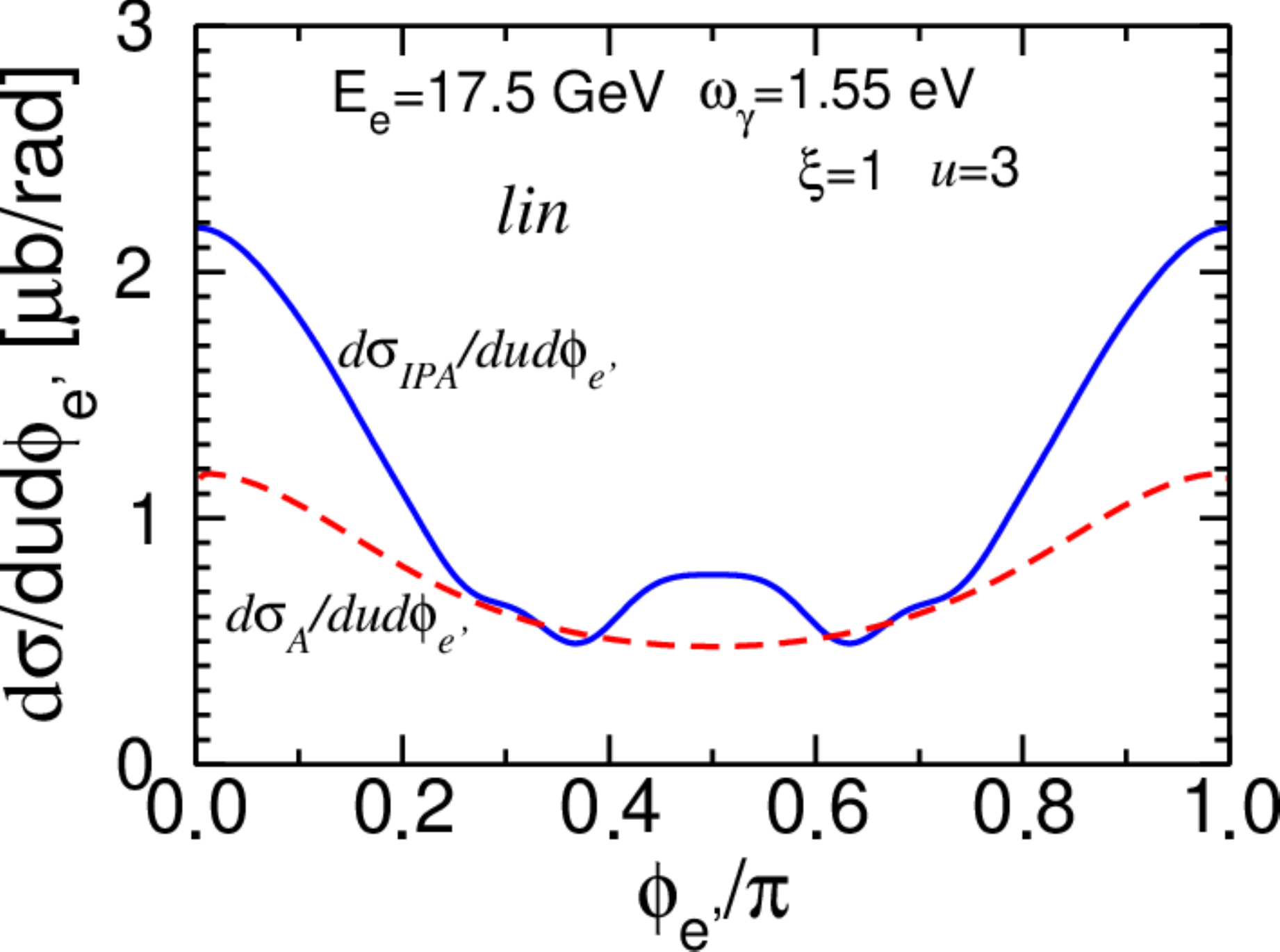}
\includegraphics[width=0.49\columnwidth]{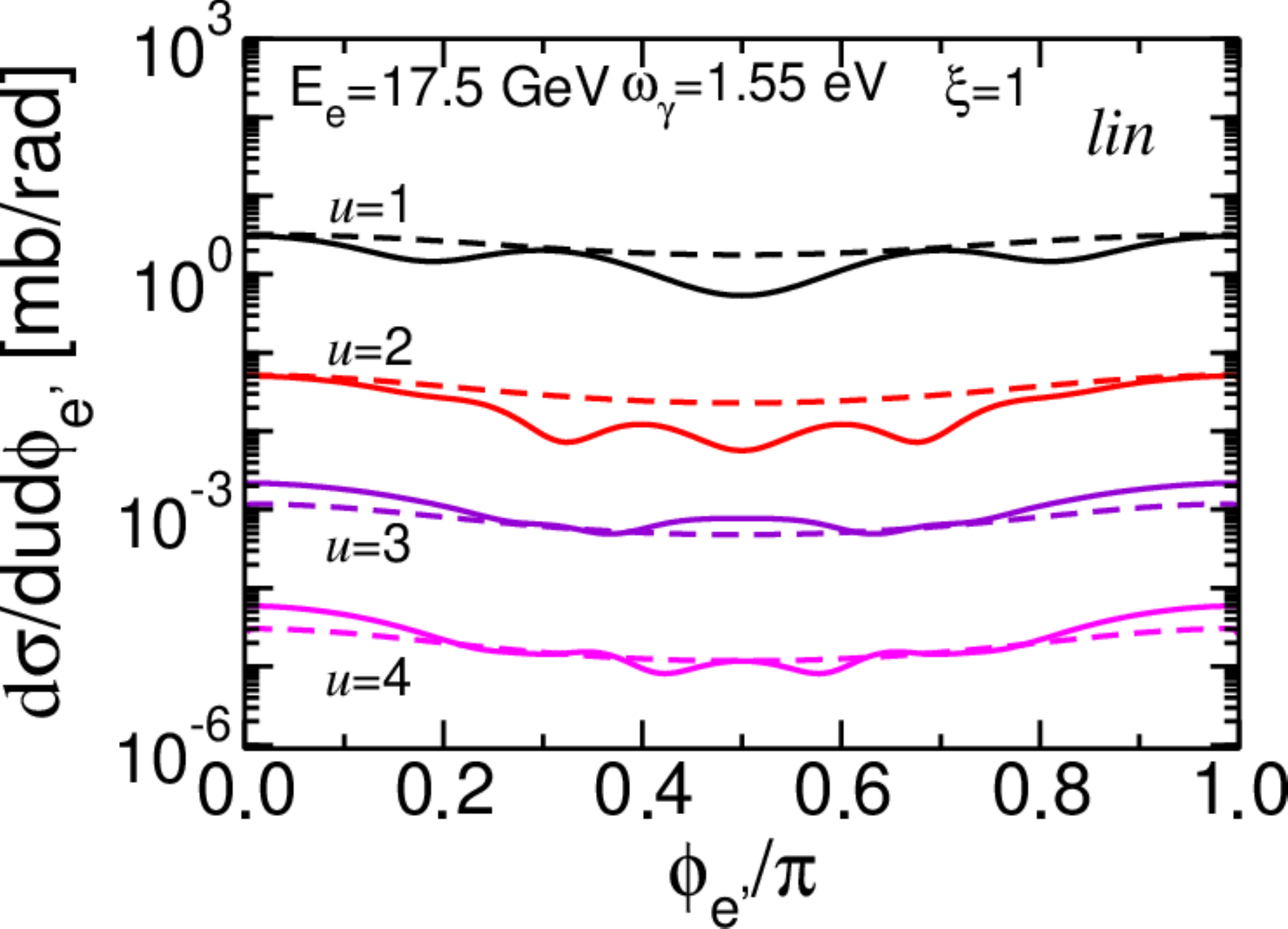}
\caption{Left panel: Invariant differential cross sections
$d^2 \sigma^{(lin)}_{IPA} / d u \, d \phi_{e'}\vert_{u = 3}$ 
according to Eq.~(\ref{sigma_IPA}) (solid curve, as in Fig.~\ref{azi},
here however in a linear scale) and 
$d^2 \sigma^{(lin)}_{IPA, large-\xi} / du \, d \phi_{e'}\vert_{u = 3}$ 
according to Eq.~(\ref{S2}) (dashed curve) for $\xi = 1$.
Right panel: 
The same as in the left panel but 
for $u = 1$, 2, 3 and 4 in $\log$ scale
with the same line style. 
For \underline{linear} polarization and $\xi = 1$.
\label{azi2}}
\end{figure}

The fine structures in the angular distribution vanish when turning
from the monochromatic model $d^2 \sigma^{(lin)}_{IPA}/du \, d \phi_{e'}$ 
to the large-$\xi$
limit, see Fig.~\ref{azi2}. In the latter case, the double-differential
cross section follows from  Eq.~(\ref{sigma_largexi})
on account of $\cos\psi=\frac{\tau}{\xi}\cot\phi_{e'}$ 
\cite{Ritus} as
\begin{eqnarray}
&&\frac{d^2 \sigma^{(lin)}_{IPA, large-\xi}}{du \, d\phi_{e'}}  =
\frac{16\alpha^2}{\pi^2\,\xi^2(k\cdot p)\,\sin^2\phi_{e'}}
\frac{1}{(1+u)^2}
\int\limits_{-\infty}^{\infty} \frac{d\tau |\tau|}{\sin\psi}  \nonumber \\
&& \times
\left( \frac{u}{2\chi\sin\psi} \right)^{\frac13}
[-\Phi^2(y) + \left( \frac{2\chi\sin\psi}{u} \right)^{\frac23}
\left(1+\frac{u^2}{2(1+u)}\right)(y\Phi^2(y) +{\Phi'}^2(y))]
 \label{S2} 
\end{eqnarray}
with $y = ( 1 + \tau^2) \left( \frac{u}{2\chi\sin\psi} \right)^{2/3}$.

Note that, in a strict head-on collision, $\phi_{\vec k'} = \pi + \phi_{e'}$,
i.e.\ the azimuthal distribution $d^2 \sigma / du \, d \phi_{e'}$
refers directly to the azimuthal photon distribution 
$d^2 \sigma / du \, d \phi_{\vec k'}$.

{\color{black}
  To summarize this part we conclude that in the IPA case the dependence
  of cross sections on the azimuthal angle of outgoing particles appears only
  for a linearly polarized laser beam.
  In the case of finite pulse duration, this dependence appears both
  for circular and linear polarizations, but becomes manifest
  more clearly for linearly polarized pulses.}

\section{q-deformed exponential}\label{qexp}

While for $\xi \approx 1$ the above spectra $d \sigma^{(i)} /du$
and $d \sigma^{(i)}_{IPA} / du$
are near to purely exponential shapes, e.g.\ for $u > 1$,
with increasing values of $\xi$ they become more convex, i.e.\ q-exponentially
deformed. Accordingly, we parameterize them by the ansatz
\begin{equation} \label{q_exp}
\frac{d \sigma_q}{d u} = \hat{\cal N} \exp_q \left(- \frac{u}{x_0} \right)
\end{equation}
in the interval $u \in [0.5, 4]$ with free normalization $\hat{\cal N}$. 
The q-exponential is defined by 
$\hat f(q, z) \equiv \exp_q(z) = [1 + (1 - q) z ]^{\frac{1}{1-q}}$; 
it obeys $\lim_{q \to 1} \exp_q (z) = \exp (z)$. The meaning of the parameters
is that of the slope at the origin, $(\partial_u \hat f / \hat f)\vert_{u=0} = - 1 / x_0$,
and the normalized curvature 
$(\hat f \partial^2_u \hat f) / (\partial_u \hat f)^2 = q$.
The series expansion
$\exp_q(-u/x_0) = \exp(- u/x_0) \left[ 1 - \frac12 (q-1) (u/x_0)^2
+ \frac{1}{24} (q-1)^2 u^3 (3 u -8)/x_0^4 + \cdots +
(q-1)^i (\cdots)  + \cdots \right]$ demonstrates the relation to a purely exponential
function, and another series expansion,
$\exp_q(-u/x_0) = 1 - (u/x_0) + \frac12 q (u/x_0)^2 + 
\frac16 q (1-2q) (u/x_0)^3 + \cdots$, also helps our understanding of the role
of the parameters $x_0$ and $q$. In particular, $q = 1$ means that a graph of
$\log f$ vs. $u$ displays a straight line.

\begin{figure}[t!] 
\includegraphics[width=0.49\columnwidth]{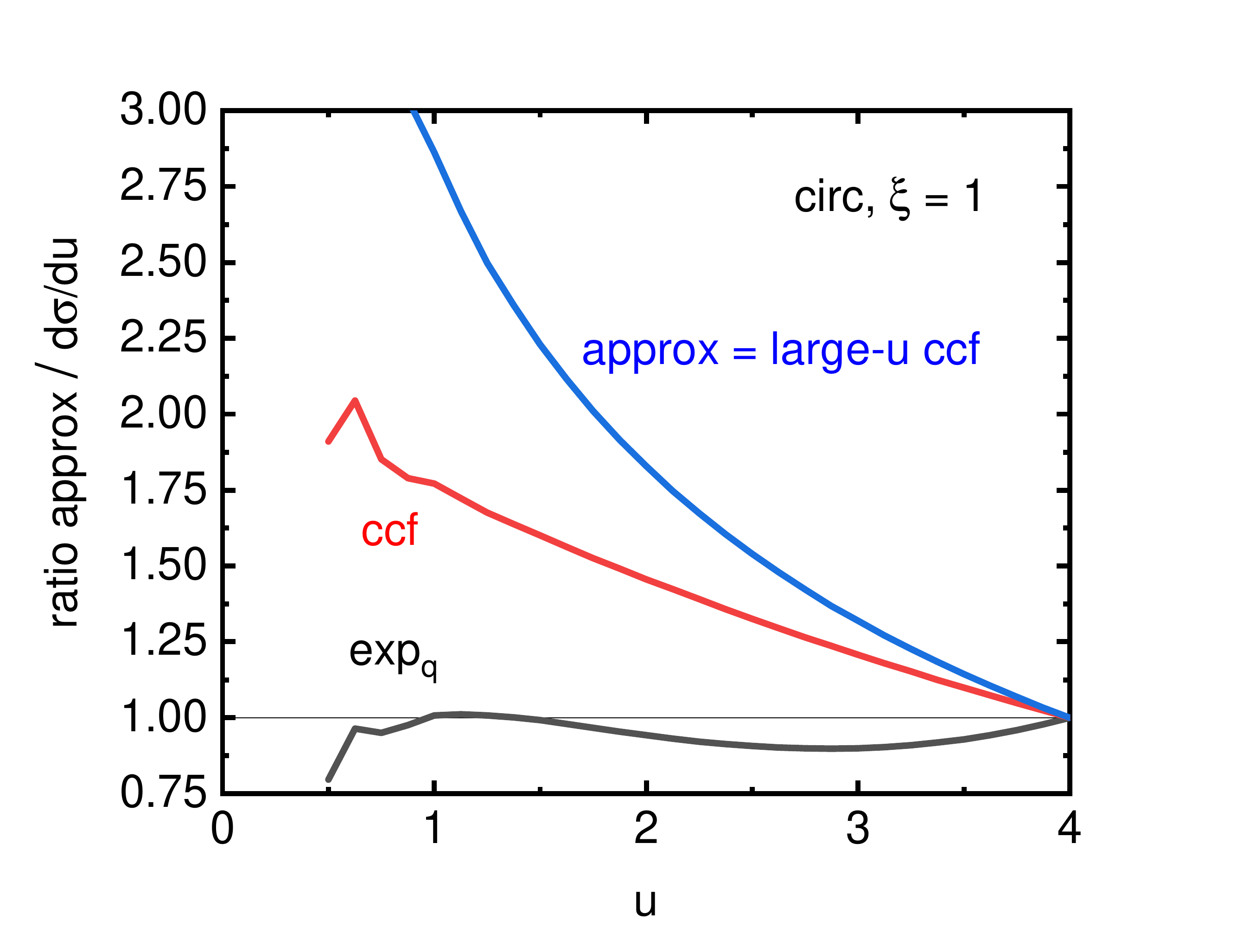}
\includegraphics[width=0.49\columnwidth]{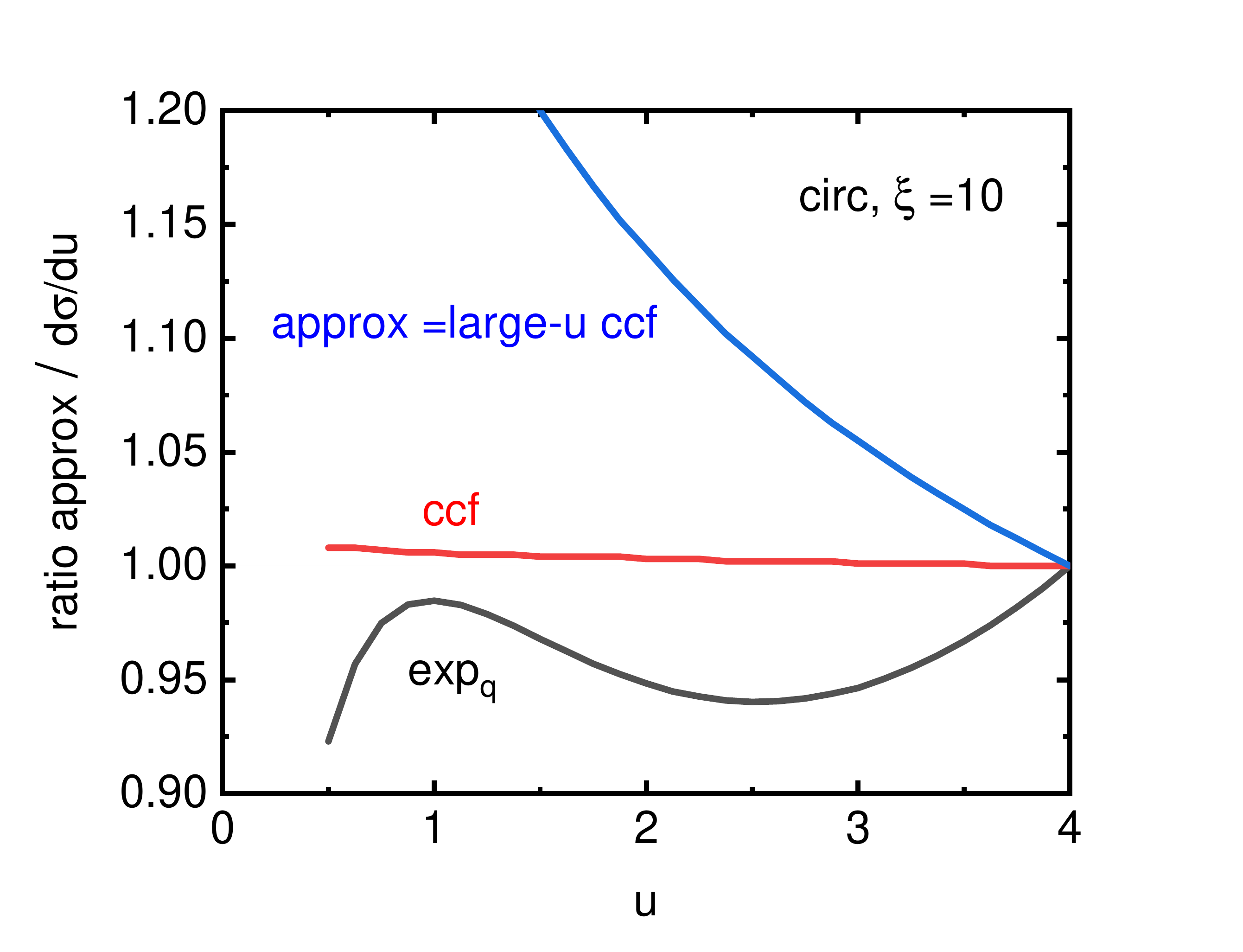}
\caption{Ratios of the q-exponential fit, Eq.~(\ref{q_exp}),
to $d \sigma^{(circ)}_{IPA} /du$ of Eq.~(\ref{sigma_IPA}) are exhibited by
black curves with label ``$\exp_q$"
for $\xi = 1$ (left panel) and $\xi =10$ (right panel). For 
illustrative purposes, the chosen normalization $\widehat{\cal N}$ is here 
such to achieve unity at $u = 4$. Analog normalizations are employed
for the approximations Eq.~(\ref{ccf1}) (red curves with label ``ccf"
for $(d \sigma_{ccf}/du)/(d \sigma^{(circ)}_{IPA} /du))$ and
Eq.~(\ref{ccf2}) (blue curves with label ``approx = large-$u$ ccf"
for $(d \sigma_{ccf, large-u}/du)/(d \sigma^{(circ)}_{IPA} /du))$ ).
For \underline{circular} polarization.
\label{fit_comparison}}
\end{figure}

The fits of $d \sigma_q/du$ describe the original spectra 
$d \sigma^{(circ)} / du$
with mean deviations typically less than 10\% and maximum local deviations
of less than $\pm 20$\%, despite the many orders of magnitude change
of $d \sigma / du$ in the considered interval of $u$. (For $\xi =1$ the
differential cross sections run over six orders of magnitude in the displayed
range of $u$.) To quantify this we exhibit in Fig.~\ref{fit_comparison}
by black curves the ratios
$(d \sigma_q/du)/(d \sigma^{(circ)}_{IPA}/du)$ by
artificially modifying the normalization $\hat{\cal N} \to \widehat{\cal N}$ 
such to get 
$\widehat{\cal N} \exp_q (- u / x_0) = d \sigma_{IPA}^{(circ)}/du$ at $u=4$.
Obviously, $\widehat{\cal N}$ is not the optimum normalization 
{\color{black} since only the sub-set $\{x_0, q\}$ is varied at fixed 
$\widehat{\cal N}$. The optimum normalization is achieved under 
unconstrained variation of $\{\hat {\cal N}, x_0, q\}$; it}
up-shifts the black curves somewhat. 
For large values of $\xi$, the constant cross field
approximation (Eq.~(\ref{ccf2}) (red curves))
describes the results of Eq.~(\ref{sigma_IPA}) 
even better than the q-exponential, see right panel of Fig.~\ref{fit_comparison},
which, however, is superior at smaller values of $\xi$, as recognizable in
the left panel. The large-$u$ approximation of the constant cross field
approximation Eq.~(\ref{ccf2}) (blue curves) turns out
to be less accurate within the considered ranges of $u$ and $\xi$. 
Nevertheless, given the huge variation of the differential cross section,
in particular for smaller values of $\xi$,  Eq.~(\ref{ccf2}) provides
a useful approximation, as pointed out above.

\begin{figure}[tb!] 
\includegraphics[width=0.49\columnwidth]{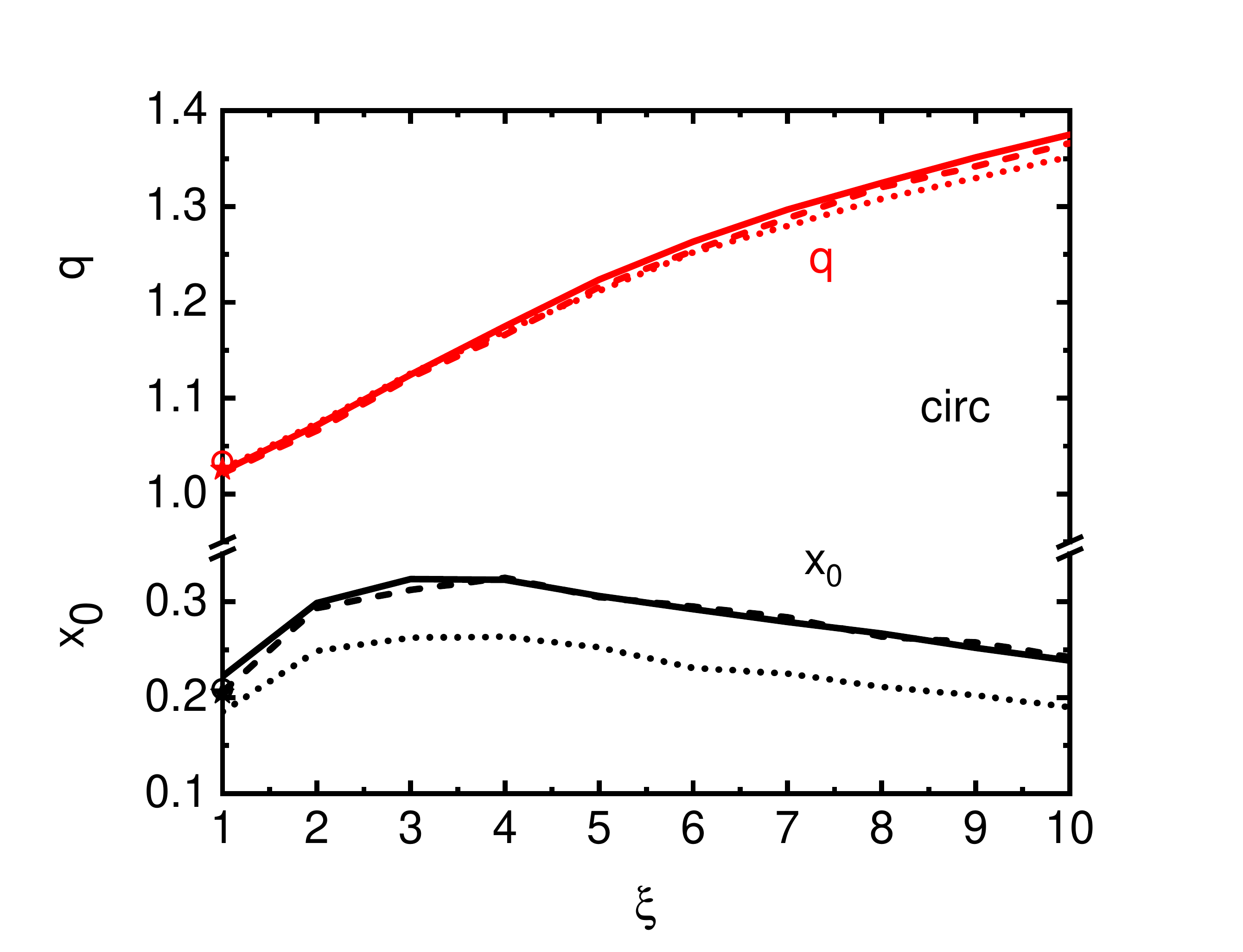}
\includegraphics[width=0.49\columnwidth]{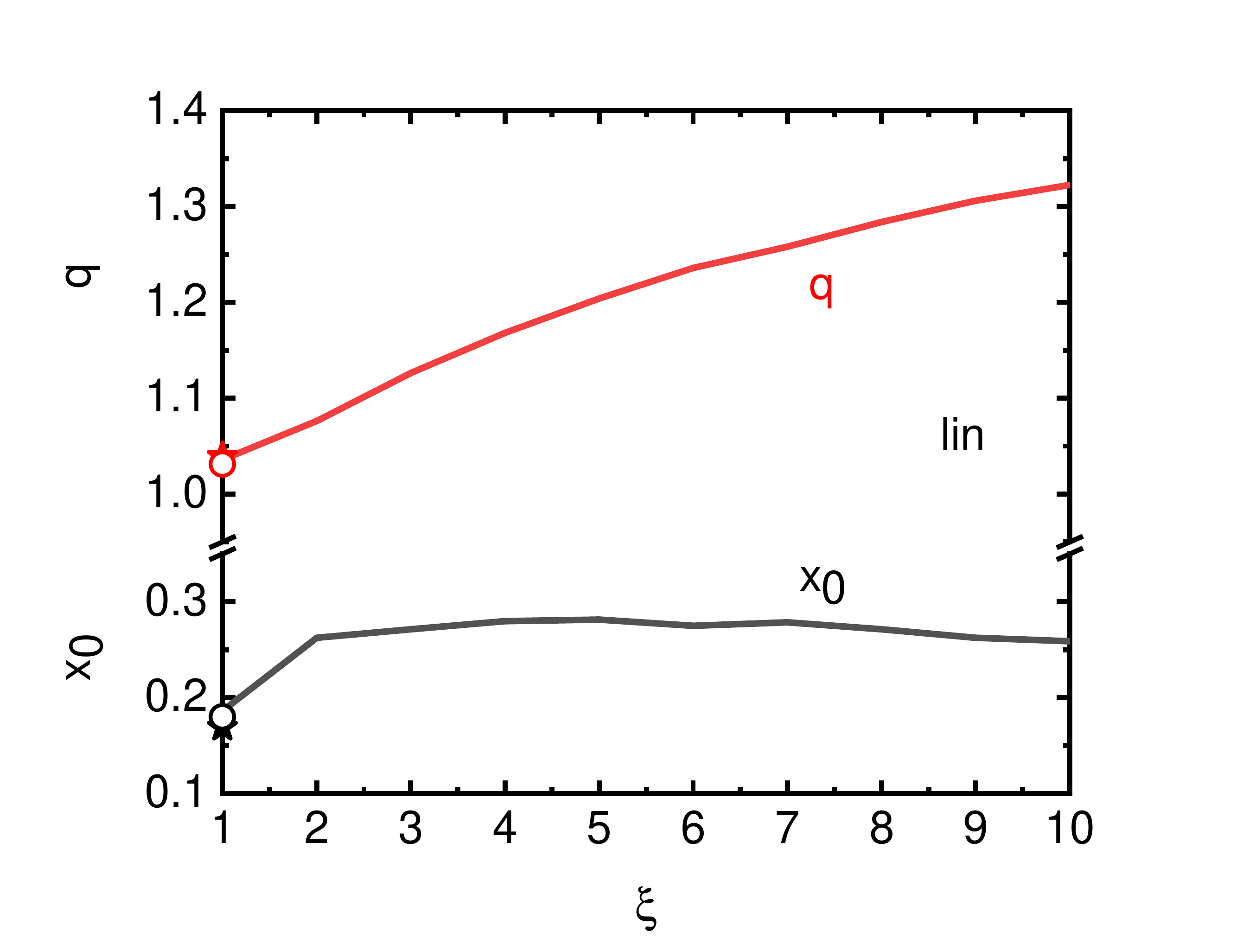}
\caption{
Fits of the cross sections $d \sigma^{(i)}_{IPA}/du$ exhibited 
in Fig.~\ref{dsigma/du_Largexi} 
by the q-exponential Eq.~(\ref{q_exp})
with parameters $x_0$ (solid black) and $q$ (solid red)
and normalization $\hat {\cal N}$ (not displayed).
The fit range is $u \in [\hat u_1, \hat u_1]$ with $\hat u_1 = 0.5$
and $\hat u_2 =4$.
{\color{black}
(This fit range uncovers the scaled $out$-photon frequencies
$\omega' / E_{e^-} = u/(1+u) \in [0.33 \cdots 0.8]$ for on-axis back-scattering.}
We mention some weak dependence of $x_0$ and $q$ and $\hat {\cal N}$
on the fit range parameters $\hat u_{1, 2}$.) 
The dashed (dotted) curves in the left panel 
depict the parameters $x_0$ and $q$ when
using Eq.~(\ref{ccf1})  ((\ref{ccf2})) with {\color{black}
  $\chi = 0.2078 \, \xi$} as input. 
The circles (asterisks) are for fits of the pulse model, $d \sigma^{(i)} /du$ 
displayed in Fig.~\ref{dsigma/du}, for $N = 1$ ($N = 10$).
Left panel: circular polarization, right panel: linear polarization.
\label{x0_q}}
\end{figure}

Coming back to the parameters of the q-exponential fits,
the resulting dependence of $x_0$ and $q$ on $\xi$ is
displayed in Fig.~\ref{x0_q} for $d \sigma^{(i)}_{IPA}/du$ as input.
Interestingly, the values of $x_0$ stay in between 0.2 and 0.35, with a maximum
at $\xi \approx 3$ for circular polarization, while for linear polarization,
$x_0$ is confined to $x_0 \in [0.16, 0.30]$. 

The parameter $q$ increases steadily from 1.02 at $\xi=1$
reaching nearly 1.4 for $\xi = 10$.
Using $d \sigma^{(i)} /du$ as input (see Fig.~\ref{dsigma/du}),
one gets the results displayed as circles ($N = 1$) and asterisks ($N = 10$).
These values do not differ noticeably from the ones obtained for the
$d \sigma^{(i)}_{IPA}/du$ input. We emphasize that such fits are in
the spirit of characterizing data, which run over many orders of magnitude,
by a few concise parameters. This is common practice, e.g.\ in particle and
relativistic heavy-ion physics (see remarks in subsection \ref{thermalization} below).

Using the approximation Eq.~(\ref{ccf1}) as input for the q-exponential fits
facilitates the dashed curves in the left panel of Fig.~\ref{x0_q}, 
which are near to the solid curves.
The large-$u$ approximation of the constant cross field
approximation, Eq.~(\ref{ccf2}), as input causes larger differences
(dotted curves), in particular for $x_0 (\xi )$. 
Nevertheless, a {\color{black}relation} emerges between
$x_0 (\chi = \xi k \cdot p /m^2 = \xi \, 0.2078)$ and the famous exponent argument 
$- 2 u / 3 \chi$ in Eq.~(\ref{ccf2}) {\color{black}at $\xi = 1$,
see the following sub-section.}

\section{Discussion}\label{discussion}

\subsection{Relation to integrated {\boldmath $u$}-differential cross sections}

Defining the integrated cross section by
$\sigma_c = \int_c^\infty du \frac{d \sigma}{du}$,
with $\sigma = \sigma_c (c \to 0)$ as total cross section,
refers to the ``cross section with cut-off $c$" considered in 
\cite{HernandezAcosta:2020agu} for $\xi = {\cal O}(1)$. Since 
$\int dz \exp_q(z) = \exp_q(z) (1+ (1-q)z)/(2 - q) +const$
one gets for the integrated cross section 
$\sigma_c = \hat{\cal N} \exp(-c/x_0) (1 - (1-q) c/x_0)/(2-q)$, when
the description (\ref{q_exp}) would apply for all values $u \ge c$. 
This implies $\sigma_c \approx \hat{\cal N} \exp(-c/x_0)$
in leading order at $q \approx 1$,
thus recovering the observation in \cite{HernandezAcosta:2020agu} that 
the partially integrated cross section displays an exponential dependence
on $1/x_0$ with $x_0 \approx 3 \chi /2$ at $\xi \approx 1$. 
We recall the relations
$\chi = \xi k \cdot p /m^2$ and $\xi = (m/\omega) ({\cal E}/{\cal E}_{crit)}$
from (\ref{I1}, \ref{I2}), 
thus $\chi = ({\cal E}/{\cal E}_{crit}) (E_{e^-}/m + \sqrt{E_{e^-}^2/m^2 - 1})$. 
At the origin of the resulting Schwinger type dependence 
$\propto \exp \left(- \frac{{\cal E}_{crit}}{{\cal E}} \frac{u}{E_{e^-}/m 
+ \sqrt{E_{e^-}^2/m^2 - 1}} \right)$
is the near-exponential shape of the differential cross section $d \sigma / du$
of the hard-photon tails of the non-linear Compton process.
The exponential $1/\chi$-dependence of the differential 
one-photon Compton cross section has been emphasized also
in \cite{Dinu:2018efz}.  

\subsection{Thermalized systems}\label{thermalization}

Thermalized systems, e.g.\ a quark-gluon plasma, with spatial extensions
smaller than the photon's mean-free path, exhibit a photon emission rate 
$\propto T^2 \exp \{ - E_\gamma / T\} \log(\tilde \alpha E_\gamma / T)$,
cf.\ \cite{McLerran:2015mda}
(here, $T$ stands for the system's temperature, $E_\gamma$ is the
photon energy and $\tilde\alpha$ denotes a system-specific parameter).
The exponential behavior reflects the thermal Boltzmann-Gibbs
distribution functions of the constituents, modified by quantum statistics. 
Otherwise, the particle 
transverse-momentum spectra observed in ultra-relativistic heavy-ion collisions
over nine orders of magnitude,
e.g.\ at the LHC 
{\color{black}(cf.\,  figure 1 in \cite{Balek:2017man} and figures 4 and 5 in \cite{Acharya:2018orn}
for examples among many others),}
maybe conveniently parameterized either by Boltzmann-Gibbs
distributions with one slope parameter (the ``temperature")
-- modified by a collective flow (resulting in J\"uttner functions and thus
modifying the exponential shapes) --
or by Tsallis distributions \cite{Tsallis:1987eu}, similar to Eq.~(\ref{q_exp}), 
which refer to non-extensive thermodynamics and statistics,
see \cite{Rath:2019cpe}.\footnote{The list of exponential distributions of quanta
emitted by special system is fairly long, ranging up to Hawking radiation
off black hole horizonts and Unruh radiation seen by an accelerated observer 
moving through the vacuum.
{\color{black} For citations, where thermal effects
have already been discussed in the literature in relation to Schwinger-type exponents in QED, cf.\ \cite{Gies:1999vb,King:2012kd,Gould:2018efv}.}}

Having these considerations in mind together with the Schwinger type behavior,
one could be tempted to {\color{black} consider}
the hard-photon emission by an electron
traversing a strong background field 
{\color{black} also} as a statistical process of shaking photons
off the field-modified vacuum by the disturbance by the electron.
The fluctuation in the related ``temperature" is directly linked to 
the non-extensive parameter $q$ and tells us about the departure 
of the system from an equilibrium state.

\subsection{Photon-energy differential cross section}

\begin{figure}[tb!] 
\includegraphics[width=0.49\columnwidth]{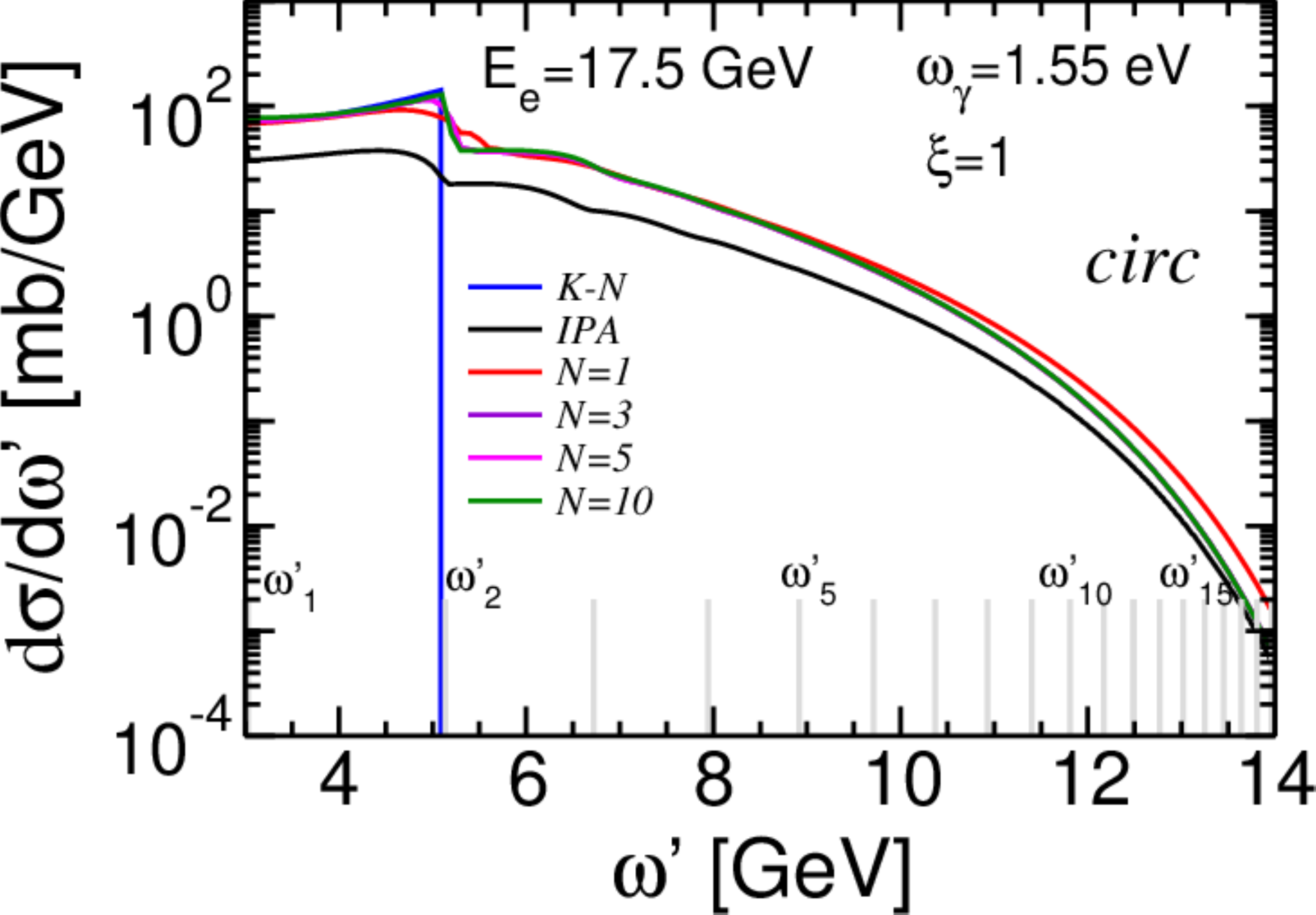}
\includegraphics[width=0.49\columnwidth]{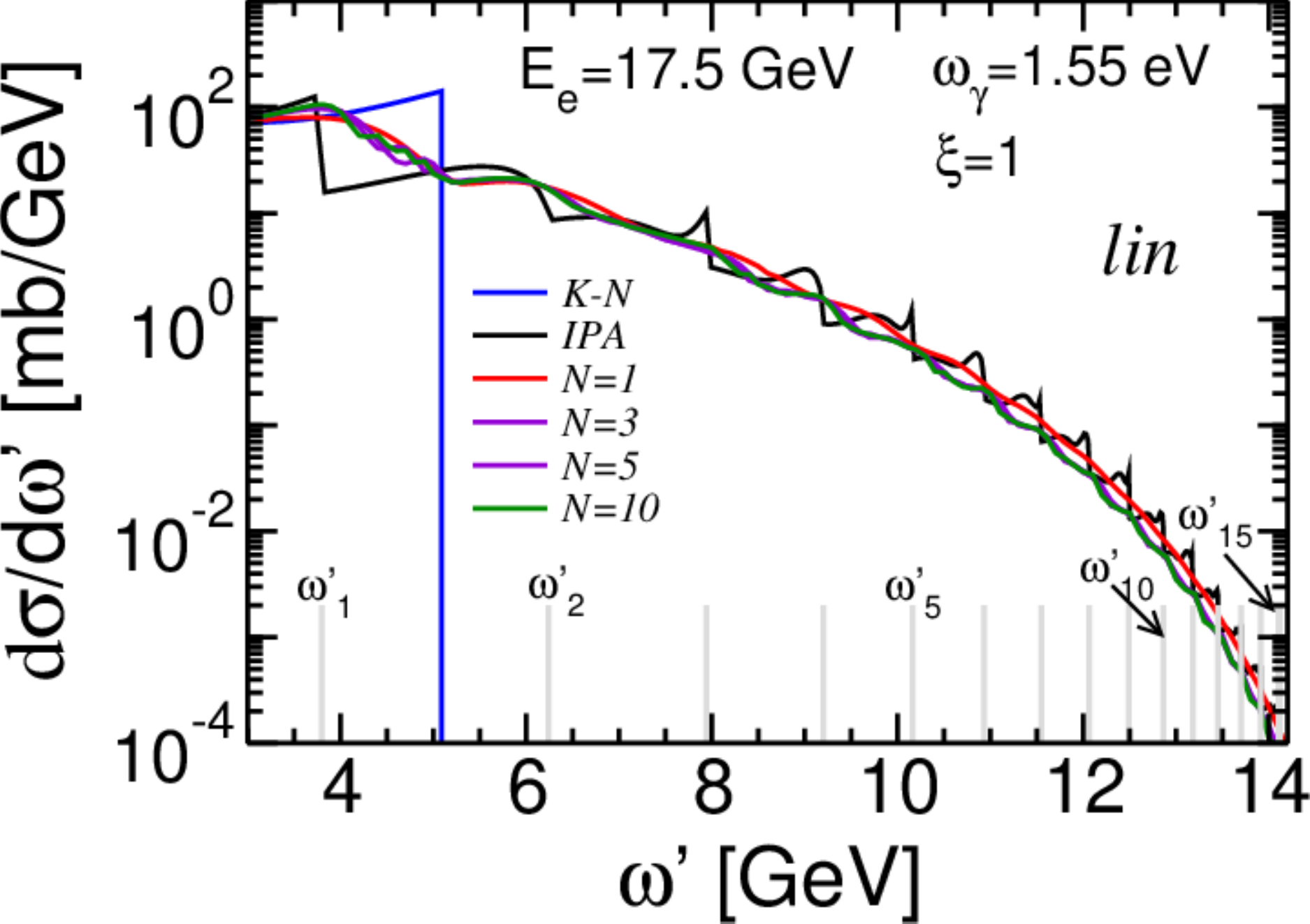}
\caption{Differential cross sections $d \sigma^{(i)}/d\omega' $ for $\xi = 1$
analog to bottom row in Fig.~\ref{dsigma/du}.
{\color{black}
The harmonic thresholds $\omega_n^\prime$ follow from inverting the above quoted
equation of $u(\nu' = \omega'/m,n)$ from energy-momentum balance.}
\label{dsigma/domega}}
\end{figure}

Previous speculations ignore the distinction of the energy/momentum
variables and the dimensionless variable $u$. In our case, the energy of the
emitted photon, $\omega' = \nu' m$, and the polar angle $\Theta'$ determine
the quantity 
$u = \frac{e^{-\zeta} \nu' (1 - \cos \Theta')}{1 - 
e^{-\zeta} \nu' (1 - \cos \Theta')}$, where the electron energy in lab.\ determines
the rapidity $\zeta$ via $E_{e^-} = m \cosh \zeta$.\footnote{
{\color{black} In this subsection we use again parameters
motivated by LUXE: 
$e^\zeta= 6.85 \times 10^4$ 
and $\nu = \omega/m = 3.033 \times 10^{-6}$}.}
Using the energy-momentum
balance in the form $u(\nu', n) = (n \nu - \nu')/(\kappa_n - n \nu + \nu')$
with $\kappa_n = n \nu - \frac12 e^\zeta + \frac12 (1 + \xi^2) e^{-\zeta}$
\cite{Harvey:2009ry} eliminates the scattering angle in favor of the harmonic
number $n$. Casting the differential cross section (\ref{sigma_IPA}) in the form
$d \sigma /du = \sum_{n=1}^\infty d \sigma_n /du$ one arrives at
$d \sigma_n/du = - (d \sigma_n / d \nu') \, \kappa_n /(1 + u(\nu', n))^2$.
For the given kinematics one can show that in leading order the relation
\begin{equation} \label{dudx}
\frac{d \sigma_n}{du}=
\frac{\vert q_z - n \omega \vert}{(1+u)^2} \frac{d \sigma_n}{d \omega'}
\approx 
\frac{\vert\kappa_{\hat n}\vert}{(1 + u (\nu', \hat n))^2}
\,\frac{d \sigma_n}{d \nu'}
\end{equation}
follows for the partial cross sections,
where $q_z = p_z - \xi^3 \omega/(2 \chi) k_z $,
$p_z = \sqrt{E_{e^-}^2 - m^2}$ and $k_z = \omega$.
This relation also holds for pulses by replacing $n \to \ell$
and $q_z \to p_z$.
Numerically, we find that any {\color{black}$n \to \hat n \ll 10^7$} is useful.
The key for the simple relations of $d \sigma / du$, $d \sigma / d \omega'$ 
and $d \sigma / d \nu'$
is $e^\zeta \gg 1$ and $\nu = \omega /m \ll 1$.
The mapping $\omega' \mapsto u$ changes the concave curves
$d \sigma / d \omega'$ as a function of $\omega'$ into convex curves
$d \sigma / d u$ as a function of $u$. The range $u \ge 1$ is beyond
the {\color{black} pronounced} harmonic structures 
-- it corresponds to $\omega' > 8.5$ GeV, {\color{black} where 
local structures can be considered as sub-leading modulations
of the gross shape, in particular for linear polarization.}
To highlight these relations we exhibit in Fig.~\ref{dsigma/domega}
the differential cross sections $d \sigma / d \omega'$ as a function of 
$\omega'$. One observes a fast decrease of the cross sections at 
$\omega' > 11$~GeV
and weak dependence on the pulse duration for $N>1$,
similar to that as for $d\sigma/du$ discussed above
in the context of the bottom panels in Fig.~\ref{dsigma/du}.
{\color{black} Note that we consider here only the case of $\xi=1$.
  At lower field intensities, e.g.\ $\xi=0.1$, the cross sections
  would display pronounced harmonic structures
  analog to that of $d\sigma/du$, exhibited in the top row of Fig.~\ref{dsigma/du}.}

\section{summary}\label{summary}

In summary we point out that the non-linear Compton process,
i.e.\ the one-photon emission by an electron moving with tens GeV
energy through an optical laser pulse of moderate intensity $\xi \gtrsim 1$,
gives rise to q-deformed exponential photon tails: 
$d \sigma / du \propto \exp_q(-u / x_0)$,
where $u = k \cdot k' /(p \cdot p - k \cdot k')$ 
is the dimensionless Ritus variable meaning the light-cone momentum-transfer
from the $in$-electron to the $out$-photon,
which is, for envisaged kinematics at LUXE and E-320,
closely related to the photon energy $\omega'$. 
For $\xi \approx 1$, the slope parameter $x_0$
is in the order of $\chi$, that is another Ritus variable, which measures the
(electric) field strength in the electron's rest system, explicitly
$\chi = ({\cal E}/{\cal E}_{crit}) (E_{e^-}/m 
+ \sqrt{(E_{e^-}/m{\color{black})}^2 - 1})$.
We emphasize the Schwinger type dependence with the
enhancement factor $E_{e^-}/m 
+ \sqrt{(E_{e^-}/m{\color{black})}^2 - 1}$ which reduces
the exponential suppression, analog to the dynamically assisted Schwinger
process with momentum space information \cite{Orthaber:2011cm}.
The (near-) exponential differential cross section results in a (near-) exponential
integrated cross section when considering only the high-energy tail
\cite{HernandezAcosta:2020agu}, 
again with Schwinger type dependence.

The high-energy photon tails with $\omega' > 1$ GeV
are accessible by detectors in development
\cite{Fleck:2020opg}, e.g.\ for the LUXE set up 
\cite{Abramowicz:2019gvx,Abramowicz:2021zja}.
The present paper tests the robustness of previous results 
\cite{HernandezAcosta:2020agu} based on the monochromatic
laser beam model with circular polarization. Focusing on the region of
$\xi \gtrsim 1$ we consider the effects of laser polarizations and laser pulse
shapes and durations as well. We find some support of the monochromatic model
even for short pulses when considering the differential spectra $d \sigma / du$.
The difference of circular and linear laser polarizations shows up most clearly
in azimuthal $out$-electron distributions, while the gross features of the
differential cross sections
$d \sigma /du$ or $d \sigma / d \omega'$ and the q-exponential parameters 
and their dependence on the laser intensity $\xi \gtrsim 1$ as well
are fairly similar. Only in the limit of a monochromatic laser beam and not too
large intensities, harmonic structures modulate noticeably the differential spectra.
The presented numerical evaluations of the essentially known formalism
may serve as benchmark for more refined approaches.
{\color{black} These should account for ponderomotive broadening effects,
beam profile simulations (cf.\ figure 6 and section V.B in
\cite{Heinzl:2009nd} for estimates of such effects)
and genuine multiple-photon emissions 
\cite{Dinu:2018efz,Blackburn:2018sfn,Dinu:2019pau}
beyond the nonlinear two-photon Compton process
\cite{Lotstedt:2009zz,Loetstedt:2009zzz,Seipt:2012tn,Mackenroth:2012rb}.} {\color{black} The q-exponential parameterization of spectra can provide
a useful interpolation tool thereby.}

Finally, we emphasize the multi-photon (up to non-perturbative) effects
which shape the photon tails, thus probing the non-linear regime of QED.
As {\color{black}another} avenue towards further developments we mention, e.g.,
extensions of the standard model of particle physics by novel degrees of
freedom, represented by dark photons or axions which may affect the
electromagnetic sector and show up as modifications of the here investigated
photon spectra and seeded subsequent processes.  

\begin{acknowledgments}
The authors gratefully acknowledge the collaboration with 
D. Seipt, T. Nousch, T. Heinzl, U. Hernandez Acosta 
and useful discussions with
A. Ilderton, K. Krajewska,  M. Marklund, C. M\"uller, S.~Rykovanov, 
and G. Torgrimsson.
A. Ringwald is thanked for explanations w.r.t.\ LUXE.
The work is supported by R.~Sauerbrey and T.~E.~Cowan w.r.t.\ the study
of fundamental QED processes for HIBEF.
\end{acknowledgments}

\end{document}